\documentclass[journal]{IEEEtran}
\usepackage{setspace}
\usepackage{amsmath}
\usepackage{amsfonts}
\usepackage{cjk}
\usepackage{indentfirst}
\usepackage{cite}
\usepackage{bm}
\usepackage{graphicx}
\usepackage{subfigure}
\usepackage{booktabs}
\usepackage{multirow}

\begin{document}

\title{Efficient and Robust Recovery of Sparse Signal and Image Using Generalized Nonconvex Regularization}

\author{Fei Wen, \textit{Member}, \textit{IEEE},
        Yuan Yang,
        Ling Pei, \textit{Member}, \textit{IEEE},
        Wenxian Yu, and
        Peilin Liu, \textit{Member}, \textit{IEEE}

\thanks{
}
\thanks{F. Wen, L. Pei, P. Liu and W. Yu are with the Department of Electronic Engineering, Shanghai Jiao Tong University, Shanghai 200240, China (e-mail: wenfei@sjtu.edu.cn; ling.pei@sjtu.edu.cn; liupeilin@sjtu.edu.cn; wxyu@sjtu.edu.cn).}
\thanks{Y. Yang is with the Air Control and Navigation Institution, Air Force Engineering University, Xian 710000, China (e-mail: yangyuankgd@126.com).}
}

\markboth{}
{Shell \MakeLowercase{\textit{et al.}}: Bare Demo of IEEEtran.cls for Journals}

\maketitle

\begin{abstract}
This work addresses the robust reconstruction problem of a sparse signal from compressed measurements. We propose a robust formulation for sparse reconstruction which employs the $\ell_1$-norm as the loss function for the residual error and utilizes a generalized nonconvex penalty for sparsity inducing. The $\ell_1$-loss is less sensitive to outliers in the measurements than the popular $\ell_2$-loss, while the nonconvex penalty has the capability of ameliorating the bias problem of the popular convex LASSO penalty and thus can yield more accurate recovery. To solve this nonconvex and nonsmooth minimization formulation efficiently, we propose a first-order algorithm based on alternating direction method of multipliers (ADMM). A smoothing strategy on the $\ell_1$-loss function has been used in deriving the new algorithm to make it convergent. Further, a sufficient condition for the convergence of the new algorithm has been provided for generalized nonconvex regularization. In comparison with several state-of-the-art algorithms, the new algorithm showed better performance in numerical experiments in recovering sparse signals and compressible images. The new algorithm scales well for large-scale problems, as often encountered in image processing.
\end{abstract}


\IEEEpeerreviewmaketitle {}

\section{Introduction}
Compressive sensing (CS) allows us to acquire sparse signals at a significantly lower rate than the classical Nyquist sampling [1]--[3], which has attracted much attention in recent years and found wide applications in radar [4], [5], communications [6], medical imaging [7], and image processing [8]--[10].
Particularly, the CS theory is relevant in some applications in image processing, such as magnetic resonant imaging (MRI) [7], image super-resolution and denoising [8], [10], [62], and hyper-spectral imaging [9].
In the CS framework, signals only need to be sampled at a rate proportional to their information content. This is based on the principle that, if a signal ${\bf{x}} \in \mathbb{R}{^n}$ is sparse, or can be sparsely represented on a basis, it can be recovered from a small number of linear measurements ${\bf{y}} = {\bf{Ax}} \in \mathbb{R}{^m}$ with $m < n$, where ${\bf{A}} \in \mathbb{R}{^{m \times n}}$ is the sensing matrix which is usually chosen to be a random matrix, such as Gaussian matrix, Bernoulli matrix, or partial Fourier matrix.
With the consideration of measurement noise, the compressed measurements can be modeled as
\begin{equation} 
{\bf{y}} = {\bf{Ax}} + {\bf{n}}
\end{equation}	
where ${\bf{n}} \in \mathbb{R}{^m}$ denotes the measurement noise. In general, the recovery of ${\bf{x}}$ from the compressed measurements ${\bf{y}}$ is an underdetermined inverse problem since $m < n$. However, the CS theory has established that if the sensing matrix ${\bf{A}}$ meets some stable embedding conditions [3], ${\bf{x}}$ can be reliably recovered via exploiting its sparse structure.

An intuitive method to reconstruct the sparse vector ${\bf{x}}$ consists in the following $\ell_0$-minimization problem
\begin{equation} 
\mathop {\min }\limits_{\bf{x}} {\left\| {\bf{x}} \right\|_0} ~~~\mathrm{subject~to}~~~ {\left\| {{\bf{Ax}} - {\bf{y}}} \right\|_2} \le \epsilon
\end{equation}	
where ${\left\| {\bf{x}} \right\|_0}$ is formally called ${\ell _0}$-norm, which counts the number of nonzero elements in the vector ${\bf{x}}$, $\epsilon>0$ constrains the strength of the residual error. Generally, the nonconvex ${\ell _0}$-minimization problem (2) is difficult to solve, known to be NP-hard. To address this problem, convex relaxation methods have been proposed, such as basis-pursuit denoising (BPDN) [11]
\begin{equation}  
\mathop {\min }\limits_{\bf{x}} {\left\| {\bf{x}} \right\|_1} ~~~\mathrm{subject~to}~~~ {\left\| {{\bf{Ax}} - {\bf{y}}} \right\|_2} \le \epsilon
\end{equation}
which relaxes the ${\ell_0}$-norm regularization into the ${\ell_1}$-norm regularization. The optimization problem (3) can be equivalently converted into an unconstrained formulation (also called LASSO [12])
\begin{equation}  
\mathop {\min }\limits_{\bf{x}} \left\{ \frac{1}{\mu}{\left\| {{\bf{Ax}} - {\bf{y}}} \right\|_2^2} + \left\| {\bf{x}} \right\|_1 \right\}
\end{equation}
where $\mu  > 0$ is a regularization parameter that balances the fidelity and sparsity of the solution. A large amount of algorithms have been developed for the BPDN and LASSO problems, such as the interior-point algorithms, homotopy [13], proximal-point algorithms [14], [15], approximate message passing (AMP) [16], [17], and alternative direction method of multipliers (ADMM) algorithm [18]. The proximal-point, AMP and ADMM algorithms enjoy much better worst-case complexity than the interior-point and Homotopy algorithms, in that the dominant computational effort is the relatively cheap matrix-vector multiplication.

The properties of ${\ell_1}$-regularization have been well studied in the context of CS. It has been demonstrated that the sparse signal ${\bf{x}}$ can be reliably recovered by ${\ell_1}$-regularized methods under some conditions of the sensing matrix ${\bf{A}}$, such as the restricted isometry property (RIP) [3], [19], incoherence condition [20], and null space property [21]. However, as a relaxation of the ${\ell_0}$-regularization, the performance of the ${\ell_1}$-regularization is limited in two aspects.  First, it would produce biased estimates for large coefficients [22]. Second, it cannot recover a signal with the least measurements [23]. As a result, the estimate given by an ${\ell_1}$-regularized method is not sparse enough in some situations. A simple example of such a case can be found in [24].

To address this limitation, many improved methods employing ${\ell _q}$-regularization have been proposed, such as the ${\ell _q}$-regularized least-squares (${\ell _q}$-LS) formulation
\begin{equation}  
\mathop {\min }\limits_{\bf{x}} \left\{ \frac{1}{\mu}{\left\| {{\bf{Ax}} - {\bf{y}}} \right\|_2^2} + \left\| {\bf{x}} \right\|_q^q \right\}
\end{equation}
with $0 \le q < 1$, where $\left\| {\bf{x}} \right\|_q^q$ is the nonconvex ${\ell _q}$ quasi-norm defined as $\left\| {\bf{x}} \right\|_q^q = \sum\nolimits_i^{} {{{\left| {{x_i}} \right|}^q}} $. Compared with ${\ell _1}$-regularization, ${\ell _q}$-regularization can yield significantly better recovery performance in many applications [25]--[33]. Under certain RIP conditions of the sensing matrix, ${\ell _q}$-regularized methods require fewer measurements to achieve reliable reconstruction than  ${\ell _1}$-regularized methods [23]. Meanwhile, the sufficient conditions for reliable reconstruction in terms of RIP for ${\ell _q}$-regularized methods are weaker than those for ${\ell _1}$-regularized methods [25], [33].

As in (2)-(5), many existing sparse recovery methods use the ${\ell _2}$-norm loss function. That is reasonable when the measurement noise is explicitly or implicitly assumed to be Gaussian distributed, since ${\ell _2}$-norm data fitting is optimal for Gaussian noise. However, the noise in practical applications often exhibits non-Gaussian properties. One important class of non-Gaussian noises arises in numerous practical situations is impulsive noise. Impulsive noise is well suited to model large outliers in measurements [34], which is frequently encountered in image processing [35]--[37]. In CS, the measurements can be corrupted by impulsive noise due to buffer overflow [38], missing data in the measurement process, bit errors in transmission [39], [40], and unreliable memory [41]. In these cases, the performance of ${\ell _2}$-loss based methods may severely degrade, since it is well-known that least-squares (LS) based methods are vulnerable to outliers.

To achieve robust sparse recovery in the presence of impulsive measurement noise, many robust methods have been developed recently. In [42], [43], the Lorentzian-norm has been used as the loss function. In [44], the ${\ell _1}$-norm has been employed as the metric for the residual error to obtain the ${\ell _1}$-regularized least-absolute (${\ell_1}$-LA) formulation
\begin{equation}  
\mathop {\min }\limits_{\bf{x}} \left\{ \frac{1}{\mu}{\left\| {{\bf{Ax}} - {\bf{y}}} \right\|_1} + \left\| {\bf{x}} \right\|_1 \right\}.
\end{equation}
Then, more computationally efficient ADMM based algorithms for this ${\ell _1}$-LA problem have been developed in [45]. Subsequently, an alternative ${\ell _1}$-norm constrained ${\ell _1}$-minimization problem has been considered in [46]. Meanwhile, the work [47] proposed a robust reconstruction method based on the Huber penalty function. ADMM and fast iterative shrinkage/thresholding algorithm (FISTA) based algorithms have been developed to efficiently solve the Huber penalty based formulation in [48].
Moreover, the ${\ell _p}$-norm loss with $0 \le p < 2$ has been considered in [49], [58].
Generally, when the measurements contain large errors or impulsive noise, these robust loss functions are able to yield dramatically better performance compared with the ${\ell _2}$-loss function. Notably, due to its simultaneous convexity and robustness against outliers, the ${\ell _1}$-loss function has found wide applications in designing robust methods, such as sparse representation based face recognition [50] and channel estimation [51].

In this paper, we consider the following $P(\cdot)$-regularized least-absolute formulation for sparse recovery
\begin{equation}  
\mathop {\min }\limits_{\bf{x}} \left\{ \frac{1}{\mu}{\left\| {{\bf{Ax}} - {\bf{y}}} \right\|_1} + P ({\bf{x}}) \right\}.
\end{equation}
where $P(\cdot)$ is a generalized nonconvex penalty for sparsity promotion, such as the hard-thresholding, smoothly clipped absolute deviation (SCAD), or $\ell_q$-norm penalty.
On the one hand, like the works [44]--[46], [50], [51], we use the ${\ell _1}$-loss function as it is less sensitive to outliers compared with the quadratic function.
It has been shown in [48] that, the ${\ell _1}$-loss based method offers considerable gain over the Huber- and Lorentzian-loss based ones.
On the other hand, unlike all of the existing robust methods [42]--[51] employing the ${\ell _1}$-regularization for sparsity inducing, we use a generalized nonconvex regularization penalty in the new formulation.
It is expected that, compared with the $\ell_1$-LA formulation (6), the new formulation retains the same robustness against outliers, while yields more accurate recovery via nonconvex regularization.

\subsection{ Contributions}
Generally, the problem (7) is difficult to solve, since in addition to the nonconvexity of the regularization term, both terms in the objective are nonsmooth. The main contributions of this work are as follows.

First, we propose an efficient first-order algorithm for the problem (7) based on ADMM.
The standard ADMM algorithm can be directly used to solve (7), but it is not convergent for a nonconvex $P(\cdot)$ as the loss term is nonsmooth.
To derive a convergent algorithm for generalized nonconvex $P(\cdot)$, a smoothing strategy of the ${\ell _1}$-loss has been adopted.
The new algorithm scales well for high-dimensional problems, as often encountered in image processing.

Second, a convergence condition of the new algorithm has been derived for a generalized nonconvex regularization penalty. Finally, we have evaluated the new algorithm via reconstruction experiments on both simulated vector-signals and images. The results showed that the new algorithm is more robust than ${\ell_2}$-loss based methods while be more accurate than ${\ell _1}$-regularization based methods.

Matlab codes for the proposed algorithm and for reproducing the results in this work are available online at https://github.com/FWen/LqLA-Sparse-Recovery.git.

\subsection{Outline and Notations}
The rest of this paper is organized as follows. Section II introduces the proximity operator for several generalized nonconvex penalty functions.
In section III, the new algorithm is presented. Section IV contains convergence analysis of the new algorithm. Section V contains experimental results.
Finally, section VI ends the paper with concluding remarks.

\textit{Notations:} For a vector $\bf{v}$, ${\rm{diag}}({\bf{v}})$ represents a diagonal matrix with diagonal elements be $\bf{v}$.
${\cal N}(0,{\sigma ^2})$ denotes a Gaussian distribution with zero-mean and variance $\sigma ^2$.
$E(\cdot)$, $\langle\cdot,\cdot\rangle$ and $(\cdot)^T$ stand for the expectation, inner product and transpose, respectively.
$\nabla f(\cdot)$ and $\partial f(\cdot)$ stand for the gradient and subdifferential of the function $f$, respectively.
$\rm{sign}(\cdot)$ denotes the sign of a quantity with $\rm{sign}(0)=0$.
${\lambda _{\max }}(\cdot)$ denotes the maximal eigenvalue of a matrix.
$I(\cdot)$ denotes the indicator function.
${\bf{I}}_n$ stands for an $n\times n$ identity matrix.
${\left\|\cdot\right\|_q}$ with $q\geq0$ denotes the $\ell_q$-norm defined as ${\left\| {\bf{x}} \right\|_q} = {(\sum\nolimits_{i = 1}^{} {{{\left| {{x_i}} \right|}^q}} )^{1/q}}$. ${\rm{dist}}({\bf{x}},S): = \inf \{ \| {{\bf{y}} - {\bf{x}}} \|_2:{\bf{y}} \in S\} $ denotes the distance from a point ${\bf{x}} \in {\mathbb{R}^n}$ to a subset $S \subset \mathbb{R}^n$. For a matrix $\bf{X}$, $\bf{X}\succeq\bf{0}$ means that it is positive-semidefinite.

\section{Proximity Operator for Sparsity Inducing Penalties}
Proximity operator plays a central role in devising efficient proximal splitting algorithms for many optimization problems, especially for nonsmooth inverse problems encountered in CS.
In this section, we introduce the proximity operator for the popular SCAD, hard-thresholding, $\ell_q$-thresholding, and soft-thresholding penalties.
For a proper and lower semicontinuous function $P(\cdot)$, its proximity operator with penalty $\eta$ ($\eta>0$) is defined as
\begin{equation} 
\mathrm{prox}_{P,\eta}({t}) = \mathrm{arg} \min\limits_{x} \left\{P({x})+\frac{\eta}{2}({x}-{t})^2 \right\}.
\end{equation}

(i) Hard-thresholding. The penalty is given by [60]
\begin{equation}
P(x) = 2 - {(|x| - \sqrt 2 )^2}I({|}x{|} < \sqrt 2 ) \notag
\end{equation}
and the corresponding thresholding function is
\begin{equation} 
{\rm{pro}}{{\rm{x}}_{P,\eta }}(t) = tI(|t| > \sqrt {2/\eta} ).
\end{equation}
Note that, the $\ell_0$-norm penalty $P(x)=|x|_0$ also results in (9).

(ii) Soft-thresholding, $P(x) = |x|$. The corresponding thresholding function is
\begin{equation} 
{\rm{pro}}{{\rm{x}}_{P,\eta }}(t) = {S_{1/\eta }}(t) = {\rm{sign}}(t)\max \left\{ {\left| t \right| - 1/\eta ,0} \right\}
\end{equation}
where ${S_\alpha}$ is well-known as the soft-thresholding/shrinkage operator.

(iii) $\ell_q$-norm ($0 < q < 1$), $P(x) = |x|^q$. In this case, the proximity operator (8) does not has a closed-form solution except for the two special cases of $q=\frac{1}{2}$ and $q=\frac{2}{3}$ [53], and it can be solved as [54]
\begin{equation} 
\mathrm{prox}_{P,\eta}({t}) =\left\{
\begin{aligned}
&0,~~~~~~~~~~~~~~~|t|< \tau \\
&{\{ 0,{\rm{sign}}({t})\beta \} },~|t|= \tau\\
&\mathrm{sign}(t)y^*,~~~~~~|t|> \tau
\end{aligned}
\right.
\end{equation}
where $\beta = {[2(1 - q)/\eta ]^{\frac{1}{{2 - q}}}}$, $\tau = \beta + q{\beta ^{q - 1}}/\eta$, ${y^*}$ is the solution of ${h}(y) = q{y^{q-1}} + \eta y - \eta | {{t}} | = 0$ over the region $(\beta ,|{t}|)$. The function $h(y)$ is convex, thus, when $|t|> \tau$, ${y^*}$ can be iteratively computed by a Newton's method.

(iv) SCAD. The penalty is given by
\begin{equation}
P(x;\lambda ) = \left\{
\begin{aligned}
&\lambda {|}x{|}, ~~~~~~~~~~~~~~~~~~{|x| < \lambda }\\
&\frac{2a\lambda {|}x{|} - {x^2} - {\lambda ^2}}{2(a - 1)}, ~{\lambda  \le {|}x{|} < a\lambda }\\
&(a + 1){\lambda ^2}/2,~~~~~~~~ {{\rm{|}}x{\rm{|}} \ge a\lambda }
\end{aligned}
\right.\notag
\end{equation}
for some $a > 2$, where $\lambda > 0$ is a threshold parameter. The corresponding thresholding function is [61]
\begin{equation}
{\rm{pro}}{{\rm{x}}_{P,\eta }}(t) = \left\{
\begin{aligned}
&{\rm{sign}}(t)\max \{ {\rm{|}}t{\rm{|}} - \lambda ,0\} ,~~{{\rm{|}}t{\rm{|}} \le 2\lambda }\\
&\frac{(a - 1)t - {\rm{sign}}(t)a\lambda}{a - 2},~~~~ {2\lambda < |t| \le a\lambda}\\
&t,~~~~~~~~~~~~~~~~~~~~~~~~~~~~{{\rm{|}}t{\rm{|}} > a\lambda }
\end{aligned}
\right..
\end{equation}

(v) Minimax concave (MC) penalty. As well as the hard, $\ell_q$, and SCAD penalties, MC can also ameliorate the bias problem of LASSO [63], and it has been widely used for penalized variable selection in high-dimensional linear regression. MC has a parametric formulation as
\begin{equation}
P(x;\lambda ) = \lambda \int_0^{|x|} {\max (1 - t/(\gamma \lambda ),0)} dt \notag
\end{equation}
with $\gamma>1$. The corresponding thresholding function is
\begin{equation}
{\rm{pro}}{{\rm{x}}_{P,\eta }}(t) = \left\{
\begin{aligned}
&0,~~~~~~~~~~~~~~~~~~~~~{|t| \le \lambda {/\eta } }\\
&\frac{{{\rm{sign}}(t)(|t| - \lambda{/\eta } )}}{{1 - 1/\gamma }},~{\lambda{/\eta }  < |t| \le \gamma \lambda{/\eta } }\\
&t,~~~~~~~~~~~~~~~~~~~~~~{|t| > \gamma \lambda {/\eta }}
\end{aligned}
 \right.. \notag
\end{equation}
For each $\lambda>0$, we can obtain a continuum of penalties and threshold operators by varying $\gamma$ in the range $(0,+\infty)$.

\section{Proposed Algorithm}
In this section, we propose an efficient algorithm to solve the $\ell_q$-LA problem (7) based on the ADMM framework. ADMM is a simple but powerful framework, which is well suited to distributed optimization and meanwhile is flexible to solve many high-dimensional optimization problems. Recently, it has found increasingly wide applications in applied statistics and machine learning [18].
ADMM has a decomposition-coordination procedure, which naturally decouples the variables and makes the global problem easy to tackle.


Specifically, with the use of an auxiliary variable ${\bf{v}}\in \mathbb{R}^m$, the formulation (7) can be rewritten as
\begin{equation} 
\mathop {\min }\limits_{{\bf{x}},{\bf{v}}} \left\{ {\frac{1}{\mu }\|{\bf{v}}\|_1 + P({\bf{x}})} \right\}~~~\textmd{subject~to}~~~{\bf{Ax}} - {\bf{y}} = {\bf{v}}.
\end{equation}
The augmented Lagrangian of the problem is
\begin{equation}
\begin{aligned}
&\mathcal{L} ({\bf{v}},{\bf{x}},{\bf{w}}) = \frac{1}{\mu }\|{\bf{v}}\|_1 + P({\bf{x}}) - \langle{\bf{w}},{\bf{Ax}} - {\bf{y}} - {\bf{v}}\rangle \\
&~~~~~~~~~~~~~~~~~~~~~~~~~~~~~~~~~~+ \frac{\rho }{2}\|{\bf{Ax}} - {\bf{y}} - {\bf{v}}\|_2^2
\end{aligned}\notag
\end{equation}
where $\mathbf{w}\in\mathbb{R}^m$ is the Lagrangian multiplier, $\rho>0$ is a penalty parameter. Then, ADMM consists of the following three steps
\begin{align} 
{{\bf{x}}^{k + 1}} & = \arg \mathop {\min }\limits_{\bf{x}} \left( {P({\bf{x}}) + \frac{\rho }{2}\left\|{\bf{Ax}} - {\bf{y}} - {{\bf{v}}^{k}} - \frac{{\bf{w}}^k}{\rho} \right\|_2^2} \right)\\
{{\bf{v}}^{k + 1}} & = \arg \mathop {\min }\limits_{\bf{v}} \left( {\frac{1}{\mu }\|{\bf{v}}\|_1 + \frac{\rho }{2}\left\|{\bf{A}}{{\bf{x}}^{k+1}} - {\bf{y}} - {\bf{v}} - \frac{{\bf{w}}^k}{\rho} \right\|_2^2} \right)\\
{{\bf{w}}^{k + 1}} &= {{\bf{w}}^k} - \rho \left({\bf{A}}{{\bf{x}}^{k + 1}} - {\bf{y}} - {{\bf{v}}^{k + 1}}\right).
\end{align}

The $\bf{x}$-update step (14) in fact solves a penalized LS problem. We use a standard trick for speeding up ADMM that solve this subproblem approximately. Specifically, let ${\bf{u}}^k = {\bf{y}} + {\bf{v}}^k + {\bf{w}}^k/\rho$, we linearize the quadratic term in the objective function of (14) at a point ${{\bf{x}}^k}$ as
\begin{equation}
\begin{aligned}
&\frac{1}{2}\left\| {{\bf{Ax}} - {{\bf{u}}^k}} \right\|_2^2\\
&\approx \frac{1}{2}\left\| {{\bf{A}}{{\bf{x}}^k} - {{\bf{u}}^k}} \right\|_2^2 + \left\langle {{\bf{x}} - {{\bf{x}}^k},{d_1}({{\bf{x}}^k})} \right\rangle  + \frac{1}{{2{\tau _1}}}\left\| {{\bf{x}} - {{\bf{x}}^k}} \right\|_2^2\\
&= \frac{1}{2}\left\| {{\bf{A}}{{\bf{x}}^k} - {{\bf{u}}^k}} \right\|_2^2 + \frac{1}{{2{\tau _1}}}\left\| {{\bf{x}} - {{\bf{x}}^k} + {\tau _1}{d_1}({{\bf{x}}^k})} \right\|_2^2\\
&~~~~~~~~~~~~~~~~~~~~~~~~~~~~~~~~~~~~~~~ - \frac{{{\tau _1}}}{2}\left\| {d_1({{\bf{x}}^k})} \right\|_2^2
\end{aligned}\notag
\end{equation}
where ${d_1}({{\bf{x}}^k}) = {{\bf{A}}^T}({\bf{A}}{{\bf{x}}^k} - {{\bf{u}}^k})$ is the gradient of the quadratic term at ${{\bf{x}}^k}$, ${\tau _1} > 0$ is a proximal parameter.
Based on this approximation, the $\bf{x}$-update step becomes easy to solve since it can be computed element-wise as the proximity operator (8)
\begin{equation} 
{{\bf{x}}^{k + 1}} = {\rm{prox}}_{P,\rho}({{\bf{b}}^k})
\end{equation}
with ${{\bf{b}}^k} = {{\bf{x}}^k} - {\tau _1}{{\bf{A}}^T}({\bf{A}}{{\bf{x}}^k} - {{\bf{u}}^k})$. As will be shown in Lemma 1 in section IV, for a generalized nonconvex penalty if $1/{\tau _1}$ is selected to be a Lipschitz constant of ${d_1}({\bf{x}})$, i.e., $1/{\tau _1} > {\lambda _{\max }}({{\bf{A}}^T}{\bf{A}})$, the augmented Lagrangian function is guaranteed nonincreasing when the ${\bf{x}}$-update step is approximately solved by (17).

The $\bf{v}$-upadte step (15) has an closed-form solution as
\begin{equation} 
{{\bf{v}}^{k + 1}} = {S_{1/(\mu \rho )}}\left( {{\bf{A}}{{\bf{x}}^{k + 1}} - {\bf{y}} - \frac{{\bf{w}}^k}{\rho} } \right).
\end{equation}

When $P(\cdot)$ is the $\ell_1$-norm penalty, the ADMM algorithm using the update steps (17), (18) and (16) reduces to the YALL1 algorithm, and it is guaranteed to converge to the global minimizer of the problem (13) if ${\tau _1} < 1/{\lambda _{\max }}({{\bf{A}}^T}{\bf{A}})$ [44].
However, for a nonconvex penalty, e.g., MC, SCAD or $\ell_q$-norm with $q<1$, the convergence of this ADMM algorithm is not guaranteed.
Empirical studies show that it always fails to converge in this case.

To develop a convergent algorithm for $q<1$, we consider a smoothed ${\ell _1}$-loss function and propose a smoothed formulation of the problem (7) as
\begin{equation} 
\mathop {\min }\limits_{\bf{x}} \left\{ {\frac{1}{\mu }{{\left\| {{\bf{Ax}} - {\bf{y}}} \right\|}_{1,\varepsilon }} + P( {\bf{x}}) } \right\}
\end{equation}
where the smoothed $\ell_1$-norm is defined as
\begin{equation} 
{\left\| {\bf{v}} \right\|_{1,\varepsilon }} = \sum\nolimits_i {{{(v_i^2 + {\varepsilon ^2})}^{\frac{1}{2}}}} \notag
\end{equation}
with $\varepsilon > 0$ be an approximation parameter. Since $\mathop {\lim }\limits_{\varepsilon  \to 0} {\left\| {\bf{v}} \right\|_{1,\varepsilon }} = {\left\| {\bf{v}} \right\|_1}$, ${\left\| {\bf{v}} \right\|_{1,\varepsilon }}$ accurately approximates ${\left\| {\bf{v}} \right\|_1}$ when $\varepsilon$ is sufficiently small. The main consideration of using such a smoothing strategy is that, the gradient of ${\left\| {\bf{v}} \right\|_{1,\varepsilon }}$ is Lipschitz continuous when $\varepsilon  > 0$, e.g., ${\nabla ^2}{\left\| {\bf{v}} \right\|_{1,\varepsilon }} \preceq \frac{1}{\varepsilon} {{\bf{I}}_n}$. As will be shown in section IV, this property is crucial for the convergence of the new algorithm in the case of a nonconvex $P(\cdot)$.

Similar to (13), the problem (19) can be equivalently expressed as
\begin{equation} 
\mathop {\min }\limits_{{\bf{x}},{\bf{v}}} \left\{ {\frac{1}{\mu }\|{\bf{v}}\|_{1,\varepsilon} + P({\bf{x}})} \right\}~~~\textmd{subject~to}~~~{\bf{Ax}} - {\bf{y}} = {\bf{v}}.
\end{equation}
The augmented Lagrangian of the problem is
\begin{equation}
\begin{aligned}
&\mathcal{L}_{\varepsilon} ({\bf{v}},{\bf{x}},{\bf{w}}) = \frac{1}{\mu }\|{\bf{v}}\|_{1,\varepsilon} + P({\bf{x}}) - \langle{\bf{w}},{\bf{Ax}} - {\bf{y}} - {\bf{v}}\rangle \\
&~~~~~~~~~~~~~~~~~~~~~~~~~~~~~~~~~~+ \frac{\rho }{2}\|{\bf{Ax}} - {\bf{y}} - {\bf{v}}\|_2^2 .
\end{aligned}
\end{equation}
Using the smoothed $\ell_1$-loss, the $\bf{v}$-update step becomes
\begin{equation}
{{\bf{v}}^{k + 1}} = \arg \mathop {\min }\limits_{\bf{v}} \left( {\frac{1}{\mu }\|{\bf{v}}\|_{1,\varepsilon} + \frac{\rho }{2}\left\|{\bf{A}}{{\bf{x}}^{k+1}} \!-\! {\bf{y}} - {\bf{v}} \!-\! \frac{{\bf{w}}^k}{\rho} \right\|_2^2} \right).
\end{equation}
As the objective function in (22) is smooth, the subproblem (22) can be solved by a standard iterative method, such as the gradient descent method, conjugate gradient method, or quasi-Newton method. However, using such an iterative method, the overall algorithm has double loops and therefore is inefficient. To improve the overall efficiency of the algorithm, we adopt the standard strategy for accelerating ADMM again, which bypasses the inner loop in this step via solving (22) approximately. Specifically, we approximate the term ${\left\| {\bf{v}} \right\|_{1,\varepsilon }}$ in the objective function of (22) by
\begin{equation} 
{\left\| {\bf{v}} \right\|_{1,\varepsilon }} \approx {\left\| {{{\bf{v}}^k}} \right\|_{1,\varepsilon }} + \left\langle {{\bf{v}} - {{\bf{v}}^k},{d_2}({{\bf{v}}^k})} \right\rangle  + \frac{1}{{2{\tau _2}}}\left\| {{\bf{v}} - {{\bf{v}}^k}} \right\|_2^2 \notag
\end{equation}
where ${d_2}({{\bf{v}}^k}) = \nabla {\left\| {{{\bf{v}}^k}} \right\|_{1,\varepsilon }}$ with ${d_2}{({{\bf{v}}^k})_i} = {v_i}{(v_i^2{\rm{ + }}{\varepsilon ^2})^{ - 1/2}}$, ${\tau _2} > 0$ is an approximation parameter. Using this linearization, the solution of the problem is explicitly given by
\begin{equation}
\begin{aligned}
{{\bf{v}}^{k + 1}} = \frac{{{\tau _2}}}{{\rho \mu {\tau _2} + 1}}\bigg[ &\frac{1}{{{\tau _2}}}{{\bf{v}}^k}- {d_2}({{\bf{v}}^k})\\
&~~~~  + \rho \mu \left( {{\bf{A}}{{\bf{x}}^{k{\rm{ + 1}}}} - {\bf{y}} - \frac{{{{\bf{w}}^k}}}{\rho }} \right) \bigg].
\end{aligned}
\end{equation}

Note that for the proposed algorithm, the dominant computational load in each iteration is matrix-vector multiplication with complexity $O(mn)$. Thus, it scales well for high-dimension problems.

\section{Convergence Analysis}

This section analyzes the convergence property of the new algorithm for a generalized nonconvex penalty.
While the convergence issue of ADMM based algorithms has been well addressed for the convex case,
there have been only a few works reported very recently on that issue for the nonconvex case [55]--[57].
The following sufficient condition for convergence is derived by using the approaches in [55], [56].
We first give the following lemmas in the proof of the main result. All the proofs are given in Appendix.

\textbf{Lemma 1.} Suppose that $P(\cdot)$ is a closed, proper, lower semicontinuous function, for any ${{\bf{x}}^k} \in \mathbb{R}{^n}$, the minimizer ${{\bf{x}}^{k+1}}$ given by (17) satisfies
\begin{equation} 
{\mathcal{L}_\varepsilon }({{\bf{v}}^k},{{\bf{x}}^{k + 1}},{{\bf{w}}^k}) \le {\mathcal{L}_\varepsilon }({{\bf{v}}^k},{{\bf{x}}^k},{{\bf{w}}^k}) - {c_{\rm{0}}}\left\| {{{\bf{x}}^{k + 1}} - {{\bf{x}}^k}} \right\|_2^2\notag
\end{equation}
where
\begin{equation} 
{c_{\rm{0}}} = \frac{\rho }{2}\left( {\frac{1}{{{\tau _1}}} - {\lambda _{\max }}({{\bf{A}}^T}{\bf{A}})} \right).\notag
\end{equation}

\textbf{Lemma 2.} For any ${{\bf{v}}^k} \in \mathbb{R}{^m}$, the minimizer ${{\bf{v}}^{k+1}}$ given by (23) satisfies
\begin{equation} 
{\mathcal{L}_\varepsilon }({{\bf{v}}^{k + 1}},{{\bf{x}}^{k{\rm{ + 1}}}},{{\bf{w}}^k}) \le {\mathcal{L}_\varepsilon }({{\bf{v}}^k},{{\bf{x}}^{k{\rm{ + 1}}}},{{\bf{w}}^k}) - {c_1}\left\| {{{\bf{v}}^{k + 1}} - {{\bf{v}}^k}} \right\|_2^2\notag
\end{equation}
where
\begin{equation} 
{c_{\rm{1}}} = \frac{1}{{\mu {\tau _2}}} + \frac{\rho }{2} - \frac{1}{{2\mu \varepsilon }}.\notag
\end{equation}

Lemma 1 and Lemma 2 establish the descent properties for the $\bf{x}$- and $\bf{v}$-subproblems, respectively.

\textbf{Lemma 3.} Suppose that $P(\cdot)$ is a closed, proper, lower semicontinuous function, let $\tilde {\mathcal{L}}({\bf{v}},{\bf{x}},{\bf{w}},\tilde{\bf{ v}}): = {\mathcal{L}_\varepsilon }({\bf{v}},{\bf{x}},{\bf{w}}) + {c_2}{\left\| {{\bf{v}} - \tilde{\bf{ v}}} \right\|_2^2}$, for $({{\bf{v}}^k},{{\bf{x}}^k},{{\bf{w}}^k})$ generated via (17), (23) and (16), if $\varepsilon > 0$ and (24) holds, then
\begin{equation} 
\begin{split}
\tilde {\mathcal{L}}({{\bf{v}}^k},{{\bf{x}}^k},{{\bf{w}}^k},{{\bf{x}}^{k - 1}}) &\ge \tilde {\mathcal{L}}({{\bf{v}}^{k + 1}},{{\bf{x}}^{k + 1}},{{\bf{w}}^{k + 1}},{{\bf{x}}^k})\\
&+ {c_{\rm{0}}}\left\| {{{\bf{x}}^{k + 1}} - {{\bf{x}}^k}} \right\|_2^2 + {c_3}{\left\| {{{\bf{v}}^{k + 1}} - {{\bf{v}}^k}} \right\|_2^2} \notag
\end{split}
\end{equation}
where ${c_2},{c_3} > 0$ are given by
\begin{equation} 
\begin{aligned}
{c_2} &= \frac{2}{{\rho {\mu ^2}}}{\left( {\frac{1}{\varepsilon } + \frac{1}{{{\tau _2}}}} \right)^2}\\
{c_3} &= \frac{1}{2}\rho  - \frac{2}{{\rho {\mu ^2}}}\left[ {\frac{2}{{{\tau _2}^2}} + \frac{2}{{{\tau _2}\varepsilon }} + \frac{1}{{{\varepsilon ^2}}}} \right] + \frac{{2\varepsilon  - {\tau _2}}}{{2\mu {\tau _2}\varepsilon }}. \notag
\end{aligned}
\end{equation}

Lemma 3 establishes the sufficient decrease property for the auxiliary function $\tilde {\mathcal{L}}$, which indicates that $\tilde {\mathcal{L}}$ is nonincreasing and thus is convergent as it is lower semicontinuous.

\textbf{Lemma 4.} Suppose that $P(\cdot)$ is a closed, proper, lower semicontinuous function, let ${{\bf{z}}^k}: = ({{\bf{v}}^k},{{\bf{x}}^k},{{\bf{w}}^k})$ with $({{\bf{v}}^k},{{\bf{x}}^k},{{\bf{w}}^k})$ generated via (17), (23) and (16), suppose that $\varepsilon > 0$, ${\tau _1} < 1/{\lambda _{\max }}({{\bf{A}}^T}{\bf{A}})$, and (24) holds, then
\begin{equation} 
\mathop {\lim }\limits_{k \to \infty } \left\| {{{\bf{z}}^{k + 1}} - {{\bf{z}}^k}} \right\|_2^2 = 0.\notag
\end{equation}
In particular, any cluster point of $\{ {{\bf{z}}^k}\}$ is a stationary point of ${\mathcal{L}_\varepsilon }$.

\textbf{Lemma 5.} Suppose that $P(\cdot)$ is a closed, proper, lower semicontinuous function, let $\tilde {\mathcal{L}}({\bf{v}},{\bf{x}},{\bf{w}},\tilde{\bf{ v}}): = {\mathcal{L}_\varepsilon }({\bf{v}},{\bf{x}},{\bf{w}}) + {c_2}{\left\| {{\bf{v}} - \tilde{\bf{ v}}} \right\|^2}$ with ${c_2}$ defined in Lemma 3, suppose that $\varepsilon > 0$, ${\tau _1} < 1/{\lambda _{\max }}({{\bf{A}}^T}{\bf{A}})$ and (24) holds, for $({{\bf{v}}^k},{{\bf{x}}^k},{{\bf{w}}^k})$ generated via (17), (23) and (16), there exists a constant ${c_4} > 0$ such that
\begin{equation} 
\begin{split}
&{\rm{dist}}(0,\partial \tilde {\mathcal{L}}({{\bf{v}}^{k{\rm{ + }}1}},{{\bf{x}}^{k{\rm{ + }}1}},{{\bf{w}}^{k{\rm{ + }}1}},{{\bf{v}}^k}))\\
& \le {c_4}\left({\left\| {{{\bf{x}}^{k + 1}} - {{\bf{x}}^k}} \right\|_2} + {\left\| {{{\bf{v}}^{k + 1}} - {{\bf{v}}^k}} \right\|_2} + {\left\| {{{\bf{v}}^k} - {{\bf{v}}^{k - 1}}} \right\|_2}\right). \notag
\end{split}
\end{equation} 

Lemma 5 establishes a subgradient lower bound for the iterate gap, which together with Lemma 4 implies that ${\rm{dist}}(0,\partial \tilde {\mathcal{L}}({{\bf{v}}^{k{\rm{ + }}1}},{{\bf{x}}^{k+1}},{{\bf{w}}^{k+1}},{{\bf{v}}^k})) \to 0$ as $k \to \infty$.

\textbf{Theorem 1.} Suppose that $P(\cdot)$ is a closed, proper, lower semicontinuous, Kurdyka-Lojasiewicz function, $\varepsilon > 0$ and ${\tau _1} < 1/{\lambda _{\max }}({{\bf{A}}^T}{\bf{A}})$, then, if
\begin{equation} 
\rho > \frac{{\sqrt {36{\varepsilon ^2} + 28{\tau _2}\varepsilon  + 17{\tau _2}^2}  + {\tau _2} - 2\varepsilon }}{{2\mu {\tau _2}\varepsilon }}
\end{equation}
the sequence $\{ ({{\bf{v}}^k},{{\bf{x}}^k},{{\bf{w}}^k})\}$ generated by the ADMM algorithm via the three steps (17), (23) and (16) converges to a stationary point of the problem (20).

In the conditions in Theorem 1, there is no restriction on the proximal parameter ${\tau _2}$. That is due to the fact that if (24) is satisfied, the sufficient decrease property of the ${\bf{v}}$-subproblem is guaranteed since ${c_1}$ in Lemma 2 is positive in this case. However, the value of ${\tau _2}$ would affect the convergence speed of the algorithm. Intensive numerical studies show that selecting a value of the same order as $\varepsilon$ for ${\tau _2}$ can yield satisfactory convergence rate.

When $\varepsilon \to 0$, the problem (20) reduces to the original problem (13) and thus the solution of (20) accurately approximates that of (13). However, from the convergence condition in Theorem 1, the penalty parameter should be chosen to be $\rho \to \infty$ in this case. Generally, with a very large value of $\rho$, the ADMM algorithm would be very slow and impractical. In practical applications, selecting a moderate value of $\varepsilon$ suffices to achieve satisfactory performance. Moreover, a standard trick to speed up the algorithm is to adopt a continuation process for the penalty parameter. Specifically, we can use a properly small starting value of the penalty parameter and gradually increase it by iteration until reaching the target value, e.g., $0 < {\rho _0} \le {\rho _1} \le  \cdots  \le {\rho _K} = {\rho _{K + 1}} = \cdots = \rho $. In this case, Theorem 1 still applies as the value of the penalty parameter turns into fixed at $\rho$ within finite iterations. Furthermore, with an initialization which is usually used for nonconvex algorithms, the new algorithm often converges quickly even in the case of a large $\rho$.

When $P(\cdot)$ is nonconvex, the formulation (19) is nonconvex and the proposed algorithm may converge to one of its many local minimizers. In this case, a good initialization is crucial for the new algorithm
to achieve satisfactory performance. Since a standard CS method (e.g., BPDN or LASSO) may break down in highly impulsive noise, it is more appropriate to employ a robust method for initialization such as ${\ell_1}$-LA (6). The ${\ell_1}$-LA problem can be solved via the ADMM update steps (17), (18) and (16), which is guaranteed to converge to the global minimizer if ${\tau _1} < 1/{\lambda _{\max }}({{\bf{A}}^T}{\bf{A}})$ [45].

\section{Numerical Experiments}

We evaluate the new method in comparison with L1LS-FISTA [15], LqLS-ADMM [55], and YALL1 [44]. L1LS-FISTA solves the ${\ell _1}$-LS problem (4). For this standard CS formulation, there exist a number of solvers, such as interior point solvers, Homotopy, ADMM [18], and FISTA [15]. All these solvers can find the global minimizer of (4) and achieve the same accuracy, but with different computational complexity. Among these solvers, ADMM and FISTA are two of the most computational efficient. LqLS-ADMM solves the ${\ell _q}$-LS formulation (5) based on ADMM. LqLS-ADMM is run with $q = 0.5$ and it is guaranteed to converge when the penalty parameter is properly chosen [55]. YALL1 solves the robust ${\ell_1}$-LA formulation (6) using an ADMM scheme. We conduct mainly two reconstruction experiments on simulated vector-signals and images, respectively.

For the proposed method, we use the $\ell_q$-norm penalty as it has a flexible parametric form that adapts to different thresholding functions while includes the hard- and soft-thresholding as special cases,
which is termed as LqLA-ADMM in the following.
It is run with ${\tau _1} = 0.99/{\lambda _{\max }}({{\bf{A}}^T}{\bf{A}})$, $\varepsilon = {10^{-3}}$, ${\tau _2} = \varepsilon $ and $\rho = \frac{3.2}{\mu \varepsilon}$, and the ${\bf{v}}$-subproblem is updated via (23). Different values of $q$, $q \in \{ 0.2,0.5,0.7\}$, are examined for LqLA-ADMM.
We use a stopping tolerance parameter of ${10^{-7}}$ for LqLA-ADMM.
Moreover, a continuation process is used for the penalty parameter as ${\rho _k} = 1.02{\rho _{k-1}}$ if ${\rho _k} < \rho$ and ${\rho _k} = \rho $ otherwise.
The two noncnvex algorithms, LqLS-ADMM and LqLA-ADMM, are initialized by the solution of YALL1.
Note that LqLA-ADMM with $q = 1$, $\varepsilon = 0$ and updated via the steps (17), (18) and (16) reduces to YALL1.

We consider two types of impulsive noise. \textit{1) Gaussian mixture noise:} we consider a typical two-term Gaussian mixture model with probability density function (pdf) given by
\begin{equation}
(1 - \xi){\cal N}(0,{\sigma ^2}) + \xi {\cal N}(0,\kappa {\sigma ^2}) \notag
\end{equation}
where $0 \le \xi < 1$ and $\kappa > 1$. This model is an approximation to Middleton's Class A noise model, where the two parameters $\xi$ and $\kappa > 1$ respectively control the ratio and the strength of outliers in the noise. In this model, the first term stands for the nominal background noise, e.g., Gaussian thermal noise, while the second term describes the impulsive behavior of the noise.
\textit{2) Symmetric $\alpha$-stable (${\rm{S}}\alpha {\rm{S}}$) noise:} except for a few known cases, the ${\rm{S}}\alpha {\rm{S}}$ distributions do not have analytical formulations. The characteristic function of a zero-location ${\rm{S}}\alpha {\rm{S}}$ distribution can be expressed as
\begin{equation}
\varphi (\omega ) = \exp \left( {ja\omega  - {\gamma ^\alpha }|\omega {|^\alpha }} \right) \notag
\end{equation}
where $0 < \alpha \le 2$ is the characteristic exponent and $\gamma > 0$ is the scale parameter or dispersion. The characteristic exponent measures the thickness of the tail of the distribution. The smaller the value of $\alpha$, the heavier the tail of the distribution and the more impulsive the noise is. When $\alpha=2$, the ${\rm{S}}\alpha {\rm{S}}$ distribution becomes the Gaussian distribution with variance $2{\gamma ^2}$. When $\alpha=1$, the ${\rm{S}}\alpha {\rm{S}}$ distribution reduces to the Cauchy distribution.

For Gaussian and Gaussian mixture noise, we use the signal-to-noise ratio (SNR) to quantify the strength of noise, which is defined by
\begin{equation}
{\rm{SNR}} = 20{\log _{10}}\left( {\frac{{{{\left\| {{\bf{A}}{{\bf{x}}^o} - E\{ {\bf{A}}{{\bf{x}}^o}\} } \right\|}_2}}}{{{{\left\| {\bf{n}} \right\|}_2}}}} \right) \notag
\end{equation}
where ${{\bf{x}}^o}$ denotes the true signal. Since an ${\rm{S}}\alpha {\rm{S}}$ distribution with $\alpha < 2$ has infinite variance, the strength of ${\rm{S}}\alpha {\rm{S}}$ noise is quantified by the dispersion $\gamma$.

All the compared methods require the selection of the regularization parameter $\mu$, which balances the fidelity and sparsity of the solution and is closely related to the performance of these methods. A popular approach is to compute the recovery along the regularization path (a set of $\mu$), and select the optimal value based on the statistical information of the noise. More specifically, for the $\ell_1$-loss based formulations, the optimal $\mu$ can be selected as the maximum value of $\mu$ such that the bound constraint on the residual is met, e.g., ${\left\| {{\bf{A\hat x}} - {\bf{y}}} \right\|_1} \le \delta $, where $\delta$ is the estimated first-order moment of the noise. The approach also applies to the new method for sufficiently small $\varepsilon$. However, this approach cannot be used in the case of ${\rm{S}}\alpha {\rm{S}}$ impulsive noise with $\alpha \le 1$, since the first-order moment of such noise is infinite. Another effective approach is to learn a value of $\mu$ via cross-validation [61]. In our experiments, to compare the methods fairly, the regularization parameter in each method is chosen by providing the best performance in terms of relative error of recovery.

\subsection{Recovery of Simulated Sparse Signals}

In the first experiment, we evaluate the compared methods using simulated sparse signals in various noise conditions. We use a simulated $K$-sparse signal of length $n=512$, in which the positions of the $K$ nonzeros are uniformly randomly chosen while the amplitude of each nonzero entry is generated according to the Gaussian distribution ${\cal N}(0,1)$. The signal is normalized to have a unit energy value. The $m \times n$ sensing matrix ${\bf{A}}$ is chosen to be an orthonormal Gaussian random matrix with $m=200$. A recovery ${\bf{\hat x}}$ is regarded as successful if the relative error satisfies
\begin{equation}
\frac{{{{\left\| {{\bf{\hat x}} - {{\bf{x}}^o}} \right\|}_2}}}{{{{\left\| {{{\bf{x}}^o}} \right\|}_2}}} \le {10^{ - 2}}. \notag
\end{equation}
Each provided result is an average over 200 independent Monte Carlo runs. Three noise conditions are considered, Gaussian noise with SNR = 30 dB, Gaussian mixture noise with $\xi=0.1$, $\kappa = 1000$ and SNR = 30 dB, and ${\rm{S}}\alpha {\rm{S}}$ noise with $\alpha = 1$ (Cauchy noise) and $\gamma = {10^{ - 4}}$.

\begin{figure}[!t]
 \centering
 \includegraphics[scale = 0.7]{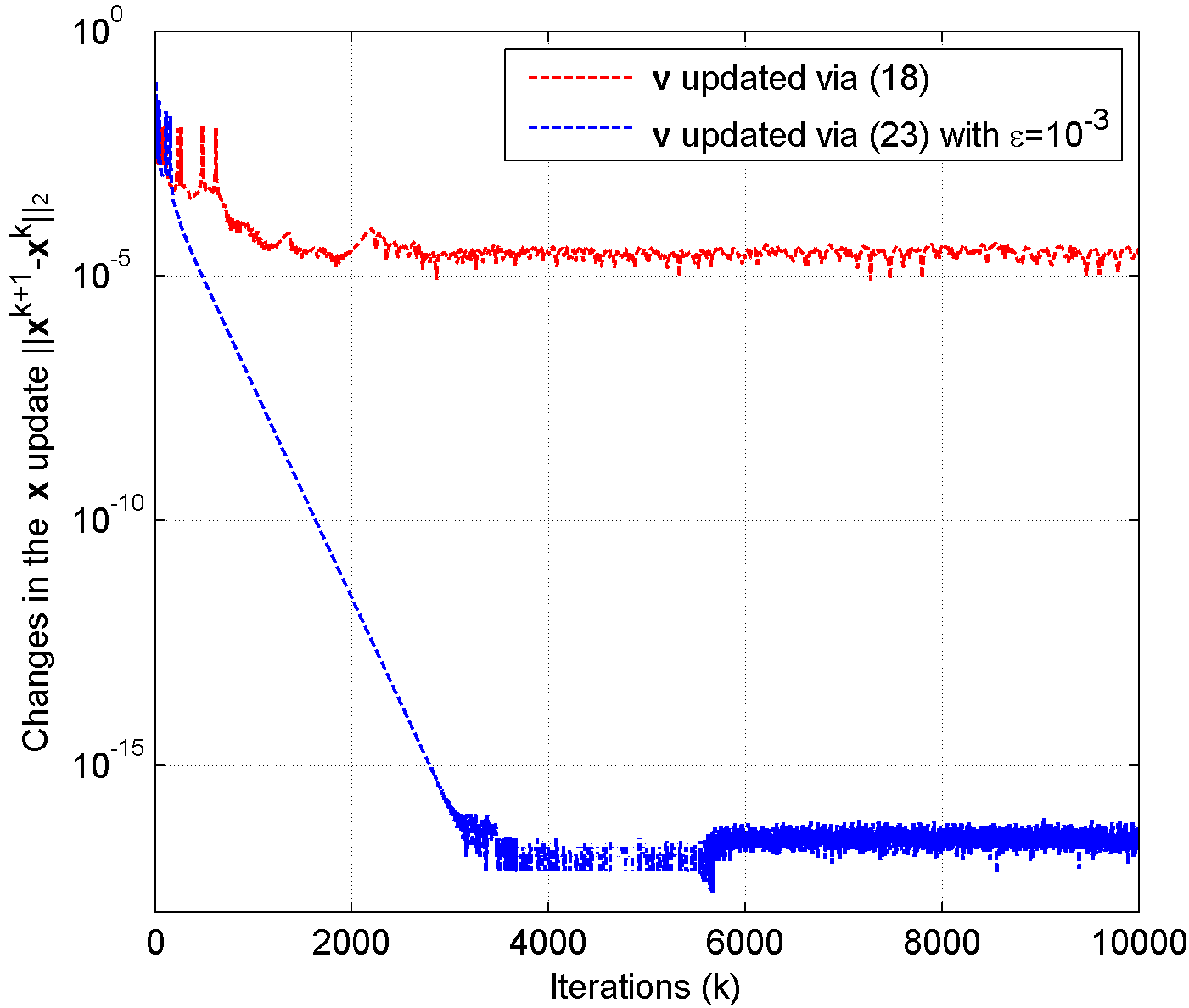}
\caption{Typical convergence behavior of LqLA-ADMM with $q=0.5$ (Gaussian mixture noise with $\xi=0.1$, $\kappa = 1000$ and SNR = 30 dB).}
 \label{figure1}
\end{figure}

\begin{figure}[!t]
\centering
\subfigure[Gaussian noise with SNR = 30 dB.]{
{\includegraphics[scale = 0.7]{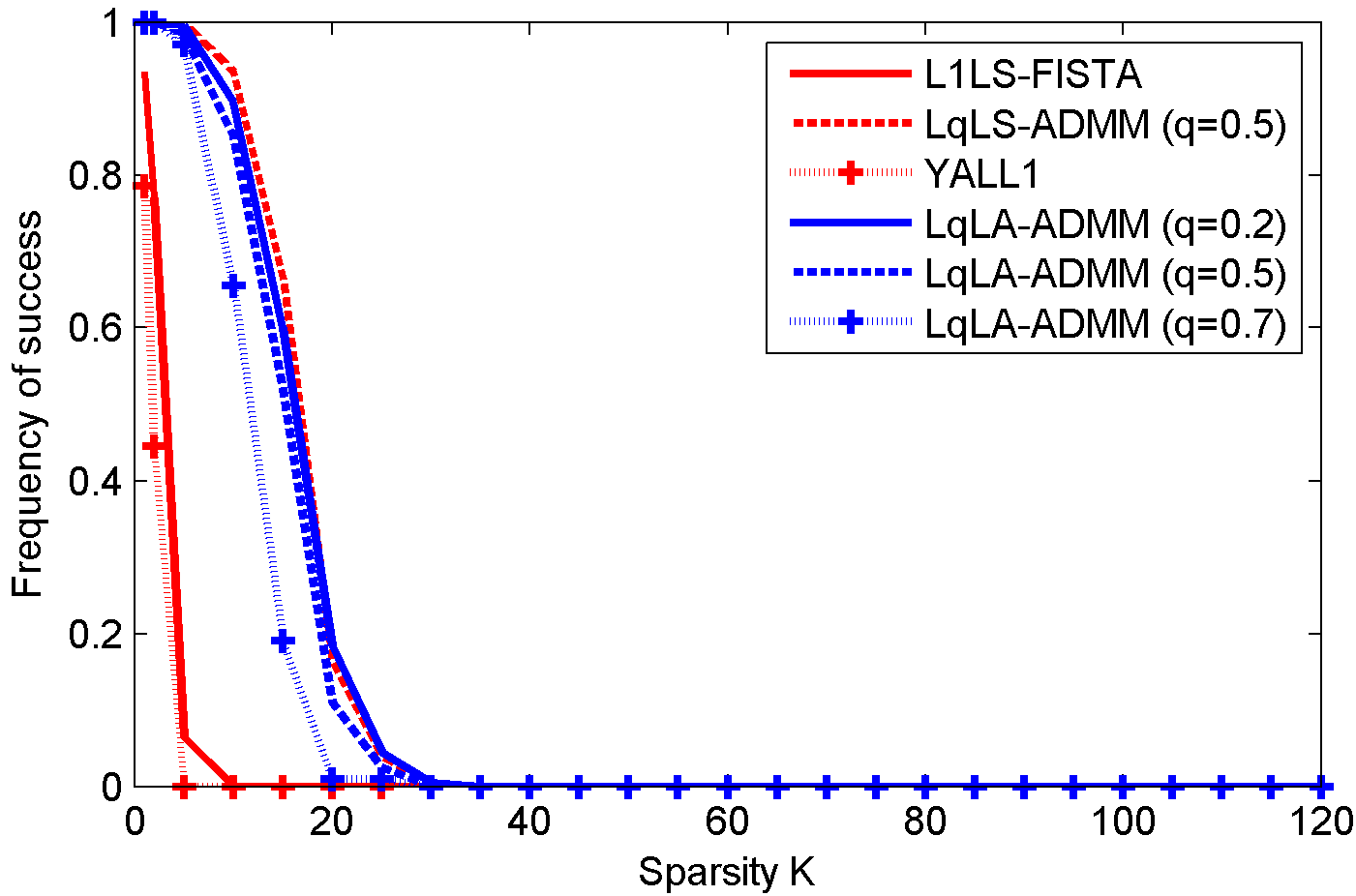}}}\\
\subfigure[Gaussian mixture noise with $\xi = 0.1$, $\kappa = 1000$ and SNR = 30 dB.]{
{\includegraphics[scale = 0.7]{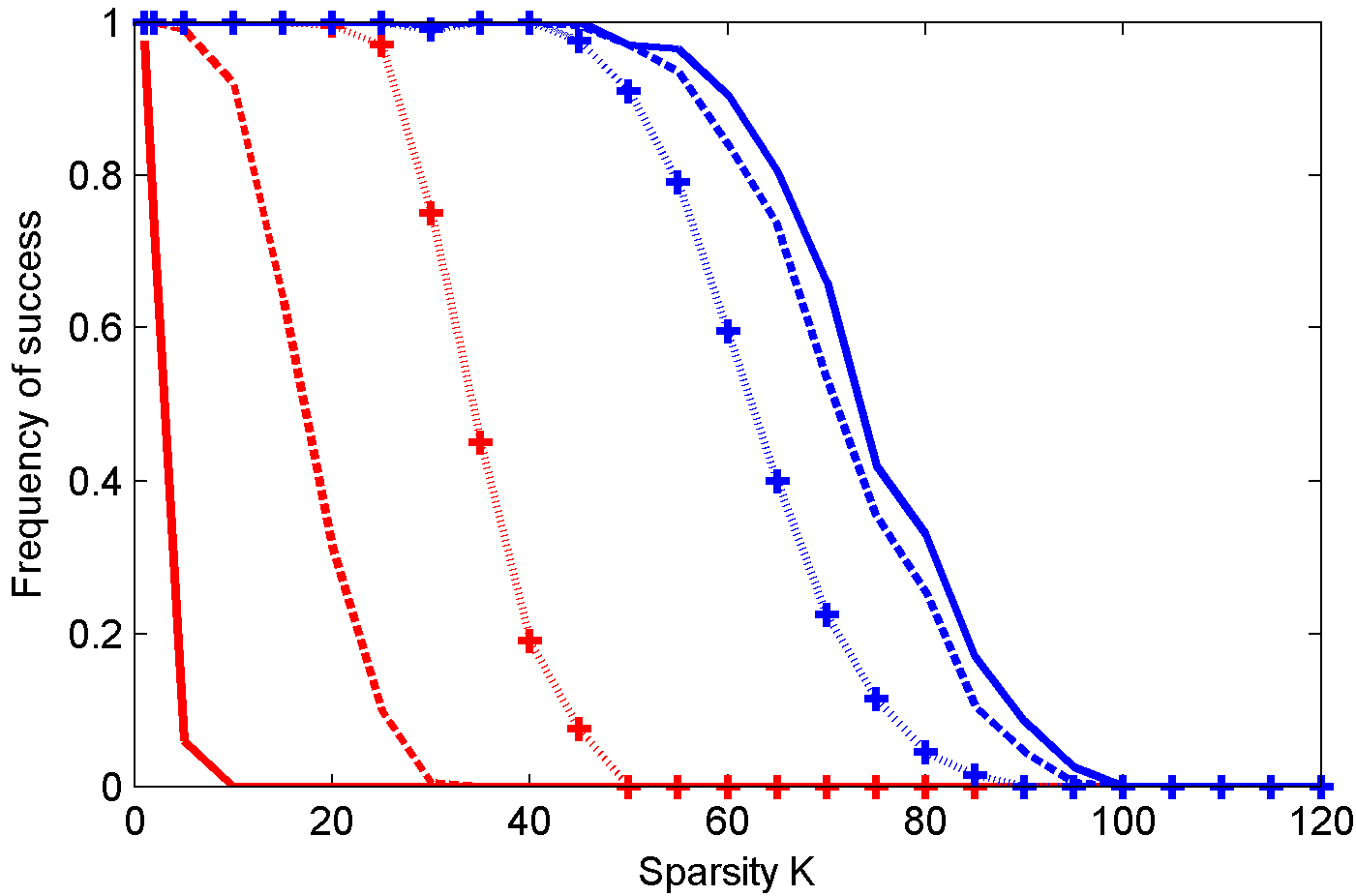}}}\\
\subfigure[$S\alpha S$ noise with $\alpha=1$ and $\gamma  = {10^{ - 4}}$.]{
{\includegraphics[scale = 0.7]{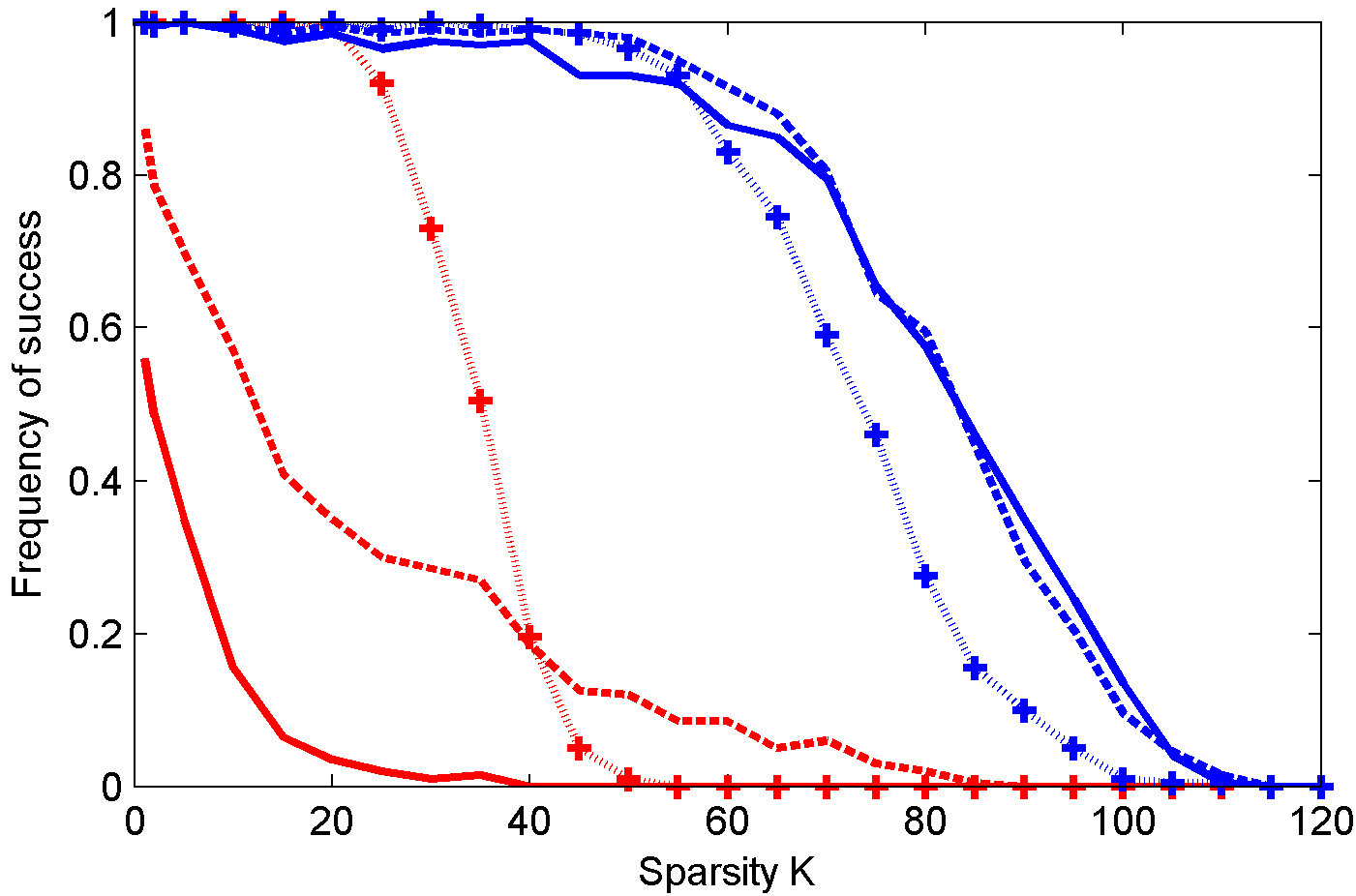}}}\\
\caption{Recovery performance versus sparsity for the compared methods in different noise conditions (a) Gaussian noise, (b) Gaussian mixture noise, (c) $S\alpha S$ noise.}
\label{figure2}
\end{figure}

Fig. 1 shows the typical convergence behavior of LqLA-ADMM with $q=0.5$ in two conditions with the $\bf{v}$-subproblem be solved by (18) and (23), respectively. The sparsity of the signal is $K=30$. It can be seen that LqLA-ADMM does not converge when the $\bf{v}$-subproblem is updated via (18).

Fig. 2 presents the successful rate of recovery of the compared algorithms versus sparsity $K$ in the three noise conditions. It is clear that in Gaussian noise, L1LS-FISTA and LqLS-ADMM respectively slightly outperform YALL1 and LqLA-ADMM. This implies that in Gaussian noise, the ${\ell _1}$-loss does not lead to considerable performance degradation relative to the ${\ell _2}$-one which is optimal in a maximum likelihood sense in this case. Moreover, LqLS-ADMM and LqLA-ADMM significantly outperform L1LS-FISTA and YALL1, which demonstrates the superiority of the ${\ell _q}$-regularization over the ${\ell _1}$-regularization.

In the two impulsive noise conditions, the ${\ell _1}$-loss based YALL1 and LqLA-ADMM algorithms outperform the ${\ell _2}$-loss based L1LS-FISTA and LqLS-ADMM algorithms in most cases. That demonstrates the robustness of ${\ell _1}$-loss against impulsive corruptions in the measurements.
Meanwhile, in impulsive noise, the advantage of ${\ell _q}$-regularization over ${\ell _1}$-regularization remains considerable.
For example, LqLA-ADMM significantly outperforms YALL1 while LqLS-ADMM significantly outperforms L1LS-FISTA. In the ${\rm{S}}\alpha {\rm{S}}$ noise condition, LqLA-ADMM can achieve a rate of successful recovery greater than $80\%$ when $K\leq 70$, while YALL1 achieves such a rate only when $K\leq 30$. Among the three tested values of $q$ ($q \in \{ 0.2,0.5,0.7\}$) for LqLA-ADMM, $q = 0.2$ and $q = 0.5$ yield better performance than $q = 0.7$.

\subsection{Recovery of Images}

\begin{figure}[!t]
 \centering
 \includegraphics[scale = 0.22]{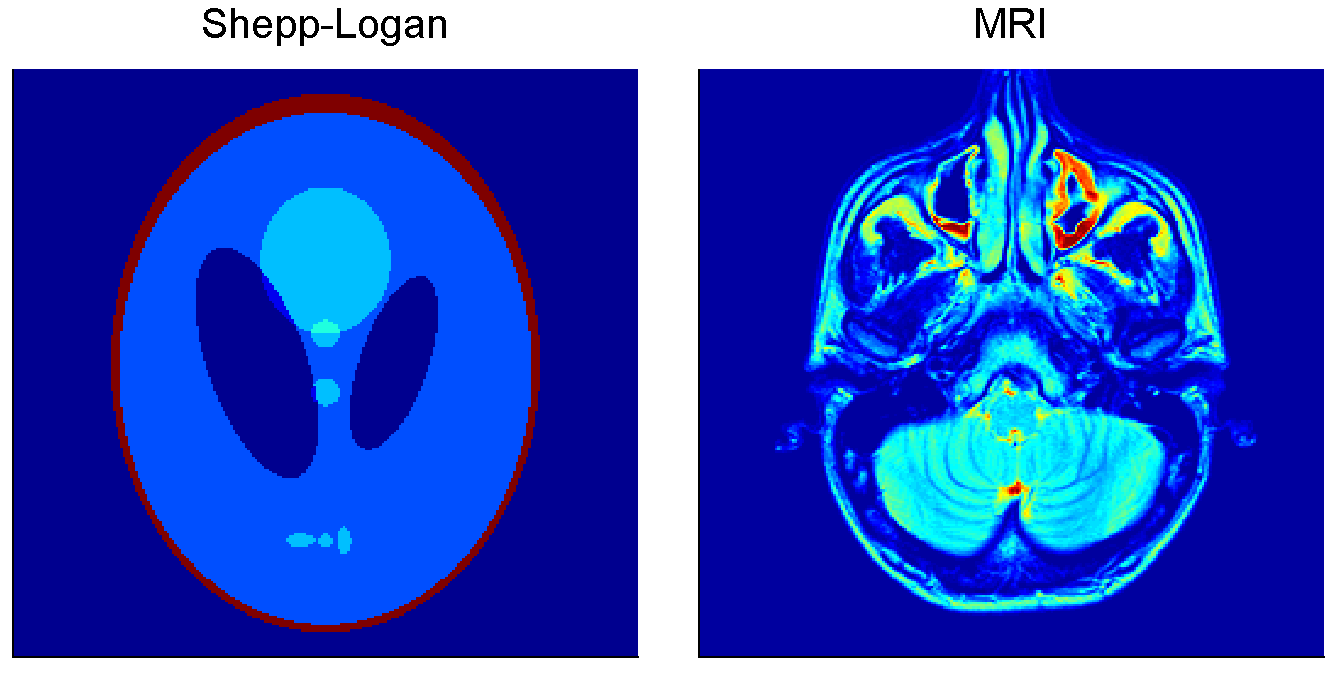}
\caption{The two $256\times256$ images used for performance evaluation.}
 \label{figure3}
\end{figure}

This experiment evaluates the algorithms on image recovery. The used images include a synthetic image, ``Shepp-Logan'', and a magnetic resonance imaging (MRI) image, as shown in Fig. 3. Each image has a size $256 \times 256$ ($n = {\rm{65536}}$), and the measurement number is set to $m = {\rm{round}}(0.4n)$. We employ a partial discrete cosine transformation (DCT) matrix as the sensing matrix ${\bf{A}}$, which is obtained by randomly selecting $m$ out of $n$ rows of the full DCT matrix. We use an implicit representation of this matrix since it is hardly explicitly available in high-dimensional conditions. Another advantage of using such a sensing matrix is that the multiplication of ${\bf{A}}$ (or ${\bf{A}}^T$) with a vector can be rapidly obtained via picking the discrete cosine transform of the vector. We use the Haar wavelets as the basis functions and consider two impulsive noise conditions, Gaussian mixture noise with $\xi=0.1$, $\kappa = 1000$ and SNR = 20 dB, and ${\rm{S}}\alpha {\rm{S}}$ noise with $\alpha = 1$ and $\gamma = {10^{ - 4}}$. The recovery performance is evaluated in terms of peak-signal noise ratio (PSNR).

\begin{figure*}[!t]
\centering
\subfigure[Shepp-Logan]{
{\includegraphics[scale = 0.25]{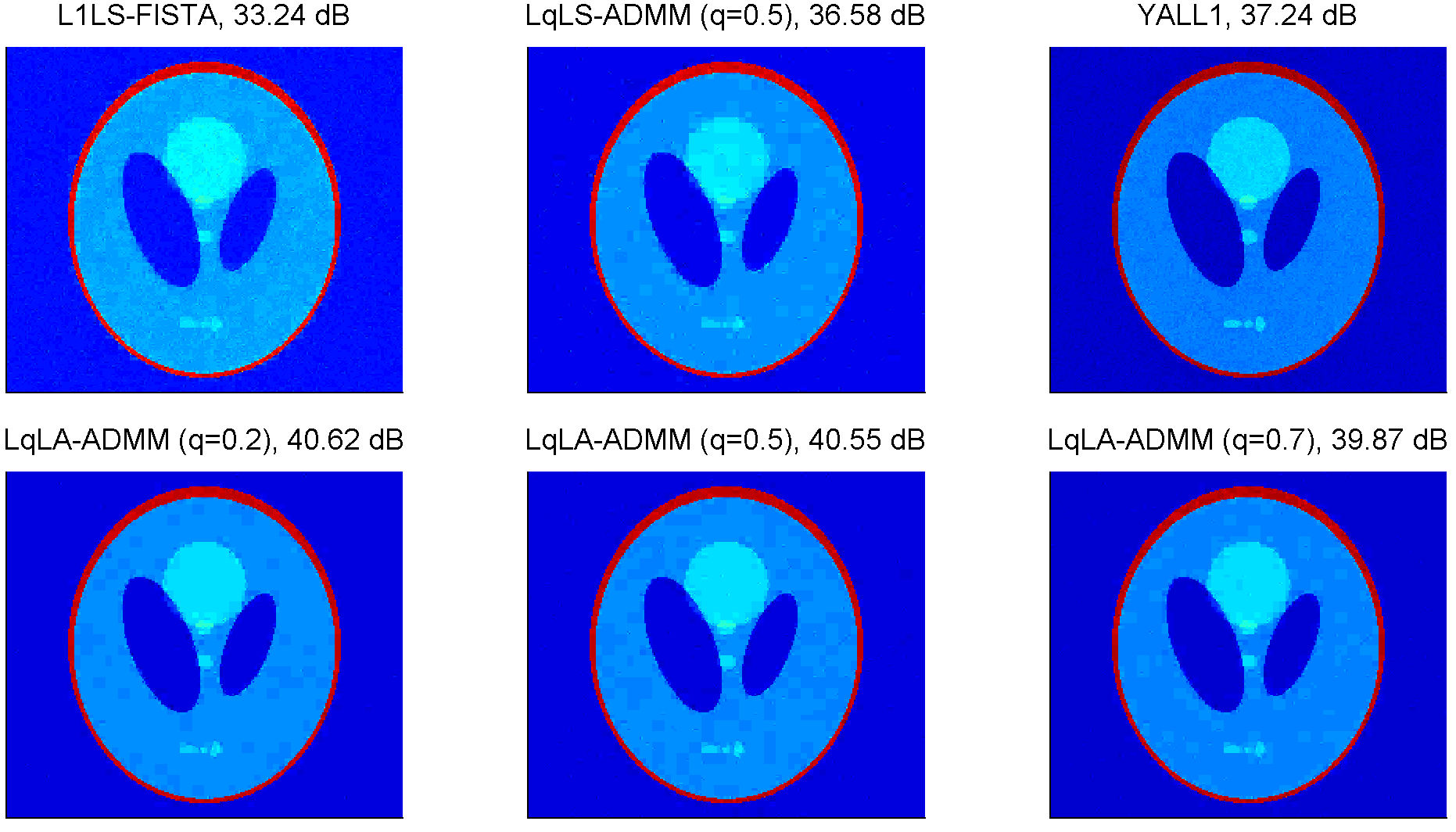}}}\\
\subfigure[MRI]{
{\includegraphics[scale = 0.25]{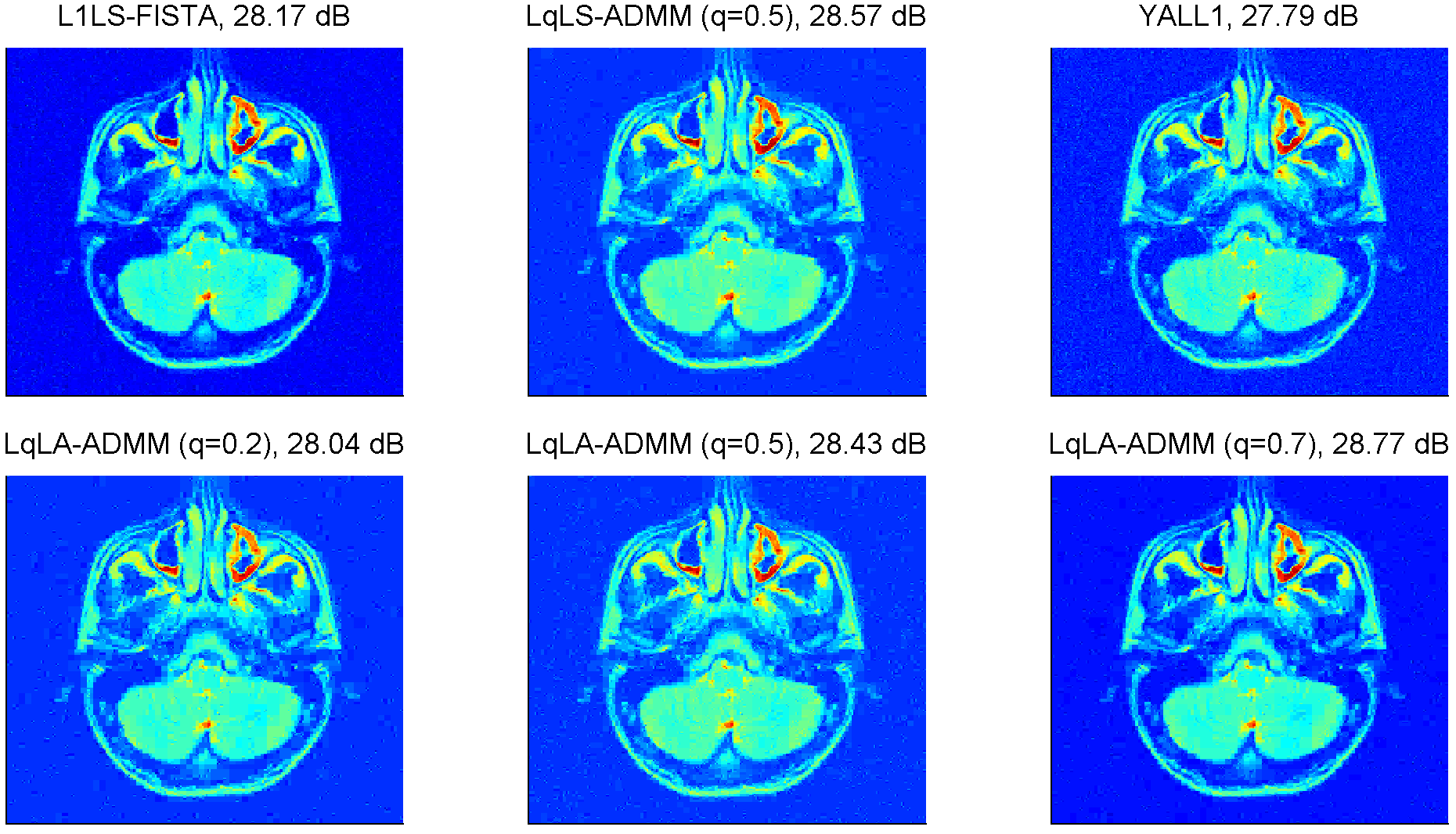}}}\\
\caption{Recovery performance of the compared methods on two $256\times256$ images in Gaussian mixture noise with $\xi = 0.1$, $\kappa = 1000$ and SNR = 20 dB.}
\label{figure4}
\end{figure*}

Fig. 4 shows the recovery performance of the compared algorithms in Gaussian mixture noise. It can be seen that each algorithm can achieve much higher PSNR in recovering the synthetic image than that in recovering the MRI image. This is due the nature that, the Haar wavelet coefficients of the synthetic image ``Shepp-Logan'' are truly sparse (approximately $3.2\%$ nonzeros), while the wavelet coefficients of a real-life image are not sparse but rather approximately follow an exponential decay, which is referred to as compressible. Moreover, LqLA-ADMM significantly outperforms the other algorithms in recovering ``Shepp-Logan'', e.g., the improvements attained by LqLA-ADMM (with $q = 0.2$) over L1LS-FISTA, LqLS-ADMM and YALL1 are 7.38, 4.04 and 3.38 dB, respectively. However, this advantage decreases in recovering the MRI image, e.g., the improvements attained by LqLA-ADMM (with $q = 0.7$) over L1LS-FISTA, LqLS-ADMM and YALL1 are 0.6, 0.2 and 0.98 dB, respectively. The results indicate that the advantage of an ${\ell _q}$-regularization based algorithm over an ${\ell _1}$-regularization based algorithm generally decreases as the compressibility of the image decreases.

Fig. 5 presents the recovery performance of the compared algorithms in the ${\rm{S}}\alpha {\rm{S}}$ noise condition. The considered ${\rm{S}}\alpha {\rm{S}}$ noise with $\alpha = 1$ contains extremely large outliers and is more impulsive than the Gaussian mixture noise. It can be seen in Fig. 5 that the ${\ell _2}$-loss based algorithms, L1LS-FISTA and LqLS-ADMM, break down, while the ${\ell _1}$-loss based algorithms, YALL1 and LqLA-ADMM, work well. LqLA-ADMM again achieves the best performance, and its advantage over YALL1 is more significant in this noise condition than that in the Gaussian mixture noise condition in recovering the MRI image. For example, in recovering the MRI image, the improvement attained by LqLA-ADMM (with $q = 0.7$) over YALL1 in Gaussian mixture noise is 0.98 dB, while that in ${\rm{S}}\alpha {\rm{S}}$ noise is 2.73 dB.

\begin{figure*}[!t]
\centering
\subfigure[Shepp-Logan]{
{\includegraphics[scale = 0.25]{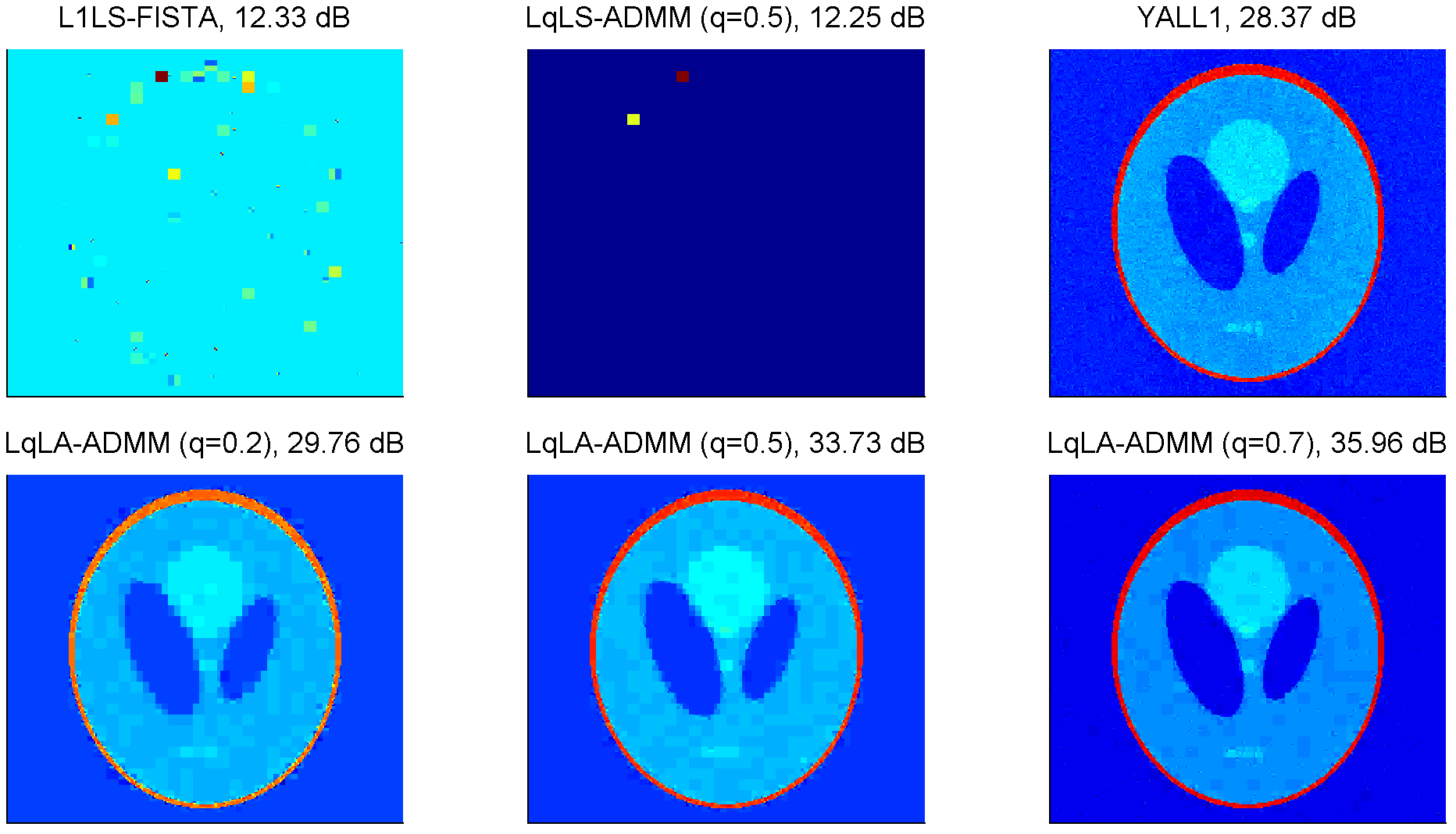}}}\\
\subfigure[MRI]{
{\includegraphics[scale = 0.25]{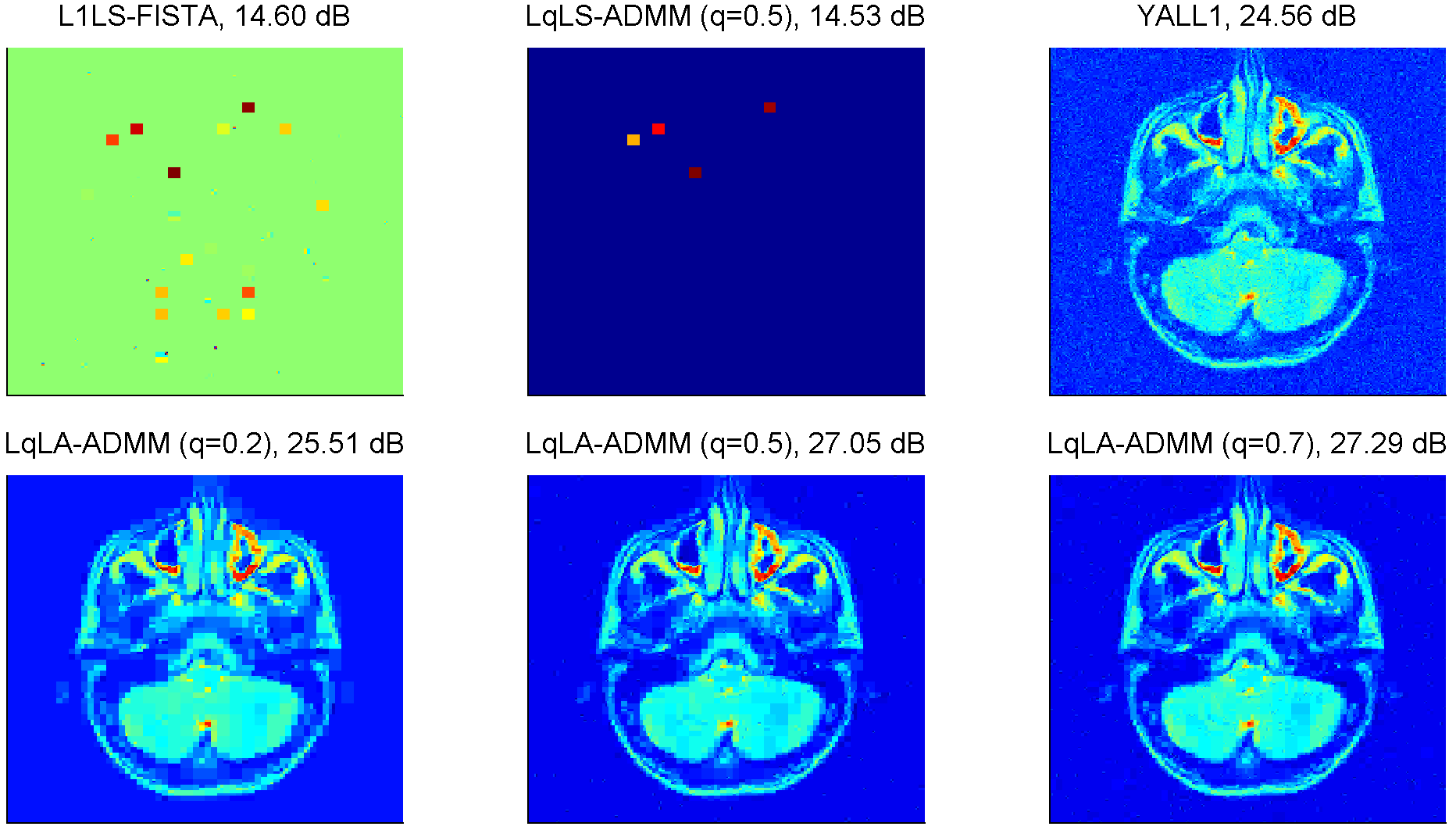}}}\\
\caption{Recovery performance of the compared methods on two $256\times256$ images in $S\alpha S$ noise with $\alpha=1$ and $\gamma  = {10^{ - 4}}$.}
\label{figure5}
\end{figure*}

Moreover, the results show that in recovering the MRI image, for LqLA-ADMM, $q=0.5$ and $q=0.7$ yield better performance than $q=0.2$, which is different form the results in the previous experiment, where $q=0.2$ and $q=0.5$ generate significantly better performance than $q=0.7$ in recovering simulated sparse signals. This is due to the nature that, real-life images are not strictly sparse as simulated sparse signals but rather compressible, e.g., with wavelet coefficients approximately follow an exponential decay.

\section{Conclusion}
This work introduced a robust formulation for sparse recovery,
which improves the ${\ell _1}$-LA formulation via replacing the ${\ell _1}$-regularization by a generalized nonconvex regularization.
A first-order algorithm based on ADMM has been developed to efficiently solve the nonconvex and nonsmooth minimization problem.
In developing the new algorithm, a smoothing strategy on the ${\ell _1}$-loss function has been used to make it convergent.
Moreover, a sufficient condition for the convergence of the new algorithm has been derived for a generalized nonconvex penalty.
Simulation results on recovering both sparse vector-valued signals and images demonstrated that, in impulsive noise, the new method offers considerable performance gain over the methods which solve the ${\ell _1}$-LS, ${\ell _q}$-LS, and ${\ell _1}$-LA formulations.

\appendices

\section{Proof of Lemma 1}
Let ${h_1}({\bf{x}}) = \frac{\rho }{2}\left\| {{\bf{Ax}} - {\bf{y}} - {{\bf{v}}^k} - {{\bf{w}}^k}{\rm{/}}\rho } \right\|_2^2$, the $\bf{x}$-subproblem in fact minimizes the following approximated objective
\begin{equation}
{Q_{{{\bf{x}}^k}}}({\bf{x}}) = P({\bf{x}}) + \left\langle {{\bf{x}} - {{\bf{x}}^k},\nabla {h_1}({{\bf{x}}^k})} \right\rangle  + \frac{\rho }{{2{\tau _1}}}\left\| {{\bf{x}} - {{\bf{x}}^k}} \right\|_2^2.\notag
\end{equation}
From the definition of ${{\bf{x}}^{k + 1}}$ as a minimizer of ${Q_{{{\bf{x}}^k}}}({\bf{x}})$, we have
\begin{equation}
\begin{split}
&{Q_{{{\bf{x}}^k}}}({{\bf{x}}^{k + 1}})\\
&= P({{\bf{x}}^{k + 1}}) + \left\langle {{{\bf{x}}^{k + 1}} - {{\bf{x}}^k},\nabla {h_1}({{\bf{x}}^k})} \right\rangle  + \frac{\rho }{{2{\tau _1}}}\left\| {{{\bf{x}}^{k + 1}} - {{\bf{x}}^k}} \right\|_2^2\\
&\le {Q_{{{\bf{x}}^k}}}({{\bf{x}}^k}) = P({{\bf{x}}^k}).
\end{split}
\end{equation}
Further, the Hessian of ${h_1}({\bf{x}})$ is
\begin{equation}
{\nabla ^2}{h_1}({\bf{x}}) = \rho {{\bf{A}}^T}{\bf{A}} \notag
\end{equation}
which implies that $\nabla {h_1}({\bf{x}})$ is $\rho {\lambda _{\max }}({{\bf{A}}^T}{\bf{A}})$-Lipschitz continuous. Thus, for any ${{\bf{x}}^k},{{\bf{x}}^{k + 1}} \in \mathbb{R}{^n}$ we have
\begin{equation}
\begin{split}
{h_1}({{\bf{x}}^{k + 1}}) &\le {h_1}({{\bf{x}}^k}) + \left\langle {{{\bf{x}}^{k + 1}} - {{\bf{x}}^k},\nabla {h_1}({{\bf{x}}^k})} \right\rangle \\
 &+ \frac{{\rho {\lambda _{\max }}({{\bf{A}}^T}{\bf{A}})}}{2}{\left\| {{{\bf{x}}^{k + 1}} - {{\bf{x}}^k}} \right\|_2^2}.
\end{split}
\end{equation}
It follows from (25) and (26) that
\begin{equation}
\begin{split}
&P({{\bf{x}}^{k + 1}}) + {h_1}({{\bf{x}}^{k + 1}})\\
& \le P({{\bf{x}}^{k + 1}}) + {h_1}({{\bf{x}}^k}) + \left\langle {{{\bf{x}}^{k + 1}} - {{\bf{x}}^k},\nabla {h_1}({{\bf{x}}^k})} \right\rangle \\
&~~~~~~~~~~~~~~~~~~~~ + \frac{{\rho {\lambda _{\max }}({{\bf{A}}^T}{\bf{A}})}}{2}{\left\| {{{\bf{x}}^{k + 1}} - {{\bf{x}}^k}} \right\|_2^2}\\
& \le P({{\bf{x}}^k}) + {h_1}({{\bf{x}}^k}) - {c_{\rm{0}}}\left\| {{{\bf{x}}^{k + 1}} - {{\bf{x}}^k}} \right\|_2^2 \notag
\end{split}
\end{equation}
which results in Lemma 1.

\section{Proof of Lemma 2}
First, the Hessian of ${\left\| {\bf{v}} \right\|_{1,\varepsilon }}$ is
\begin{equation}
{\nabla ^2}{\left\| {\bf{v}} \right\|_{1,\varepsilon }} = {\varepsilon ^2}{\rm{diag}}\{ {(v_1^2{\rm{ + }}{\varepsilon ^2})^{ - \frac{3}{2}}}, \cdots ,{(v_N^2{\rm{ + }}{\varepsilon ^2})^{ - \frac{3}{2}}}\}  \preceq \frac{1}{\varepsilon }{{\bf{I}}_n}
\end{equation}
which implies that $\nabla {\left\| {\bf{v}} \right\|_{1,\varepsilon }}$ is $\frac{1}{\varepsilon} $-Lipschitz continuous, thus, for any ${{\bf{v}}^k},{{\bf{v}}^{k + 1}} \in \mathbb{R}{^m}$ we have
\begin{equation}
\begin{split}
{\left\| {{{\bf{v}}^{k + 1}}} \right\|_{1,\varepsilon }} \le {\left\| {{{\bf{v}}^k}} \right\|_{1,\varepsilon }} &+ \langle {{{\bf{v}}^{k + 1}} - {{\bf{v}}^k},\nabla {{\| {{{\bf{v}}^k}} \|}_{1,\varepsilon }}} \rangle\\
& + \frac{1}{{2\varepsilon }}\left\| {{{\bf{v}}^{k + 1}} - {{\bf{v}}^k}} \right\|_2^2.
\end{split}
\end{equation}
Let ${h_{\rm{2}}}({\bf{v}}) = \frac{\rho }{2}\left\| {{\bf{A}}{{\bf{x}}^{k{\rm{ + 1}}}} - {\bf{y}} - {\bf{v}} - {{\bf{w}}^k}{\rm{/}}\rho } \right\|_2^2$, the $\bf{v}$-subproblem actually minimizes the following approximated objective
\begin{equation}
{G_{{{\bf{v}}^k}}}({\bf{v}}) = \frac{1}{\mu }\left\langle {{\bf{v}} - {{\bf{v}}^k},\nabla {{\left\| {{{\bf{v}}^k}} \right\|}_{1,\varepsilon }}} \right\rangle  + \frac{1}{{2\mu {\tau _2}}}\left\| {{\bf{v}} - {{\bf{v}}^k}} \right\|_2^2 + {h_{\rm{2}}}({\bf{v}}).
\end{equation}%
Since ${G_{{{\bf{v}}^k}}}({\bf{v}})$ is $(\frac{1}{\mu {\tau _2}} + \rho )$-strongly convex, for any ${{\bf{v}}^k} \in \mathbb{R}{^m}$ we have
\begin{equation}
\begin{split}
{G_{{{\bf{v}}^k}}}({{\bf{v}}^k}) &\ge {G_{{{\bf{v}}^k}}}({{\bf{v}}^{k + 1}}) + \left\langle {{{\bf{v}}^k} - {{\bf{v}}^{k + 1}},\nabla {G_{{{\bf{v}}^k}}}({{\bf{v}}^{k + 1}})} \right\rangle\\
&~~~~~~~~~~~~~~ + \frac{1}{2}\left( {\frac{1}{{\mu {\tau _2}}} + \rho } \right)\left\| {{{\bf{v}}^k} - {{\bf{v}}^{k + 1}}} \right\|_2^2.
\end{split}
\end{equation}
From the definition of ${{\bf{v}}^{k + 1}}$ as a minimizer of ${G_{{{\bf{v}}^k}}}({\bf{v}})$, we have $\nabla {G_{{{\bf{v}}^k}}}({{\bf{v}}^{k + 1}}) = 0$. Further, since ${G_{{{\bf{v}}^k}}}({{\bf{v}}^k}) = {h_{\rm{2}}}({{\bf{v}}^k})$, it follows from (29) and (30) that
\begin{equation}
\begin{split}
&\frac{1}{\mu }\left\langle {{{\bf{v}}^{k + 1}} - {{\bf{v}}^k},\nabla {{\left\| {{{\bf{v}}^k}} \right\|}_{1,\varepsilon }}} \right\rangle  + {h_{\rm{2}}}({{\bf{v}}^{k + 1}})\\
& \le {h_{\rm{2}}}({{\bf{v}}^k}) - \left( {\frac{1}{{\mu {\tau _2}}} + \frac{\rho }{2}} \right)\left\| {{{\bf{v}}^{k + 1}} - {{\bf{v}}^k}} \right\|_2^2 \notag
\end{split}
\end{equation}
which together with (28) yields
\begin{equation}
\begin{split}
&\frac{1}{\mu }{\left\| {{{\bf{v}}^{k + 1}}} \right\|_{1,\varepsilon }} + {h_{\rm{2}}}({{\bf{v}}^{k + 1}})\\
& \le \frac{1}{\mu }{\left\| {{{\bf{v}}^k}} \right\|_{1,\varepsilon }} + {h_{\rm{2}}}({{\bf{v}}^k}) - {c_{\rm{1}}}\left\| {{{\bf{v}}^{k + 1}} - {{\bf{v}}^k}} \right\|_2^2 \notag
\end{split}
\end{equation}
which finally results in Lemma 2.

\section{Proof of Lemma 3}
First, we show that the changes in the dual iterates can be bounded by the changes in the primal iterates. Observe that the approximated $\bf{v}$-subproblem actually minimizes the objective ${G_{{{\bf{v}}^k}}}({\bf{v}})$ given in (29), whose minimizer ${{\bf{v}}^{k + 1}}$ satisfies
\begin{equation}
\begin{split}
\nabla {\left\| {{{\bf{v}}^k}} \right\|_{1,\varepsilon }} &+ \frac{1}{{{\tau _2}}}\left({{\bf{v}}^{k + 1}} - {{\bf{v}}^k}\right)\\
& + \mu \rho \left({\bf{A}}{{\bf{x}}^{k{\rm{ + 1}}}} - {\bf{y}} - {{\bf{v}}^{k + 1}} - {{{{\bf{w}}^k}} / \rho }\right) = {\bf{0}}.
\end{split}
\end{equation}
Substituting (16) into (31) yields
\begin{equation}
{{\bf{w}}^{k + 1}} = \frac{1}{\mu }\nabla {\left\| {{{\bf{v}}^k}} \right\|_{1,\varepsilon }} + \frac{1}{{\mu {\tau _2}}}({{\bf{v}}^{k + 1}} - {{\bf{v}}^k}).
\end{equation}
Then, it follows that
\begin{equation}
\begin{split}
&\left\| {{{\bf{w}}^{k + 1}} - {{\bf{w}}^k}} \right\|_2^2\\
&\le \frac{1}{{{\mu ^2}}}\bigg(
{\left\| {\nabla {{\left\| {{{\bf{v}}^k}} \right\|}_{1,\varepsilon }} - \nabla {{\left\| {{{\bf{v}}^{k - 1}}} \right\|}_{1,\varepsilon }}} \right\|_2} + \frac{1}{{{\tau _2}}}{\left\| {{{\bf{v}}^{k + 1}} - {{\bf{v}}^k}} \right\|_2}\\
&~~~~~~~~~~~~~~~~~~~~~~~~~~~~~~~~~~~~~~~~ + \frac{1}{{{\tau _2}}}{\left\| {{{\bf{v}}^k} - {{\bf{v}}^{k - 1}}} \right\|_2}\bigg)^2\\
&\le \frac{1}{{{\mu ^2}}}{\left( {\frac{1}{{{\tau _2}}}{{\left\| {{{\bf{v}}^{k + 1}} - {{\bf{v}}^k}} \right\|}_2} + \left( {\frac{1}{\varepsilon } + \frac{1}{{{\tau _2}}}} \right){{\left\| {{{\bf{v}}^k} - {{\bf{v}}^{k - 1}}} \right\|}_2}} \right)^2}\\
&\le \frac{2}{{{\mu ^2}\tau _2^2}}\left\| {{{\bf{v}}^{k + 1}} - {{\bf{v}}^k}} \right\|_2^2 + \frac{2}{{{\mu ^2}}}{\left( {\frac{1}{\varepsilon } + \frac{1}{{{\tau _2}}}} \right)^2}\left\| {{{\bf{v}}^k} - {{\bf{v}}^{k - 1}}} \right\|_2^2
\end{split}
\end{equation}
where the second inequality follows from (27).

From (16) and the definition of ${\mathcal{L}_\varepsilon }$, we have
\begin{equation}
\begin{split}
{{\mathcal{L}}_\varepsilon }({{\bf{v}}^{k + 1}},{{\bf{x}}^{k + 1}},{{\bf{w}}^{k + 1}}) &- {{\mathcal{L}}_\varepsilon }({{\bf{v}}^{k + 1}},{{\bf{x}}^{k + 1}},{{\bf{w}}^k})\\
& = \frac{1}{\rho }{\left\| {{{\bf{w}}^{k + 1}} - {{\bf{w}}^k}} \right\|_2^2}.
\end{split}
\end{equation}
Then, with the use of (33), it follows from Lemma 1, Lemma 2 and (34) that
\begin{equation}
\begin{split}
&{{\mathcal{L}}_\varepsilon }({{\bf{v}}^{k + 1}},{{\bf{x}}^{k + 1}},{{\bf{w}}^{k + 1}}) - {{\mathcal{L}}_\varepsilon }({{\bf{v}}^k},{{\bf{x}}^k},{{\bf{w}}^k})\\
& \le  - {c_{\rm{0}}}\left\| {{{\bf{x}}^{k + 1}} - {{\bf{x}}^k}} \right\|_2^2 - \left( {{c_{\rm{1}}} - \frac{2}{{\rho {\mu ^2}\tau _2^2}}} \right)\left\| {{{\bf{v}}^{k + 1}} - {{\bf{v}}^k}} \right\|_2^2\\
&~~~~~~~~~~~~~~~~~~~~~~~~~+ \frac{2}{{\rho {\mu ^2}}}{\left( {\frac{1}{\varepsilon } + \frac{1}{{{\tau _2}}}} \right)^2}\left\| {{{\bf{v}}^k} - {{\bf{v}}^{k - 1}}} \right\|_2^2 \notag
\end{split}
\end{equation}
which consequently results in Lemma 3, where ${c_3}$ is positive when (24) holds. Moreover, it is easy to see that, when (24) is satisfied, ${c_{\rm{1}}}$ in Lemma 2 is also positive, which implies the sufficient decrease of ${{\mathcal{L}}_\varepsilon }$ by the ${\bf{v}}$-subproblem updated via (23).

\section{Proof of Lemma 4}
First, we show the sequence $\{ {{\bf{z}}^k}\} $ generated via (17), (23) and (16) is bounded. From (32), we have
\begin{equation}
\begin{split}
&\left\| {{{\bf{w}}^k}} \right\|_2^2 \le \frac{1}{\mu }{\left( {{{\left\| {\nabla {{\left\| {{{\bf{v}}^{k - 1}}} \right\|}_{1,\varepsilon }}} \right\|}_2} + \frac{1}{{{\tau _2}}}{{\left\| {{{\bf{v}}^k} - {{\bf{v}}^{k - 1}}} \right\|}_2}} \right)^2}\\
& \le \frac{2}{{{\mu ^2}}}\left\| {\nabla {{\left\| {{{\bf{v}}^{k - 1}}} \right\|}_{1,\varepsilon }}} \right\|_2^2 + \frac{2}{{{\mu ^2}\tau _2^2}}\left\| {{{\bf{v}}^k} - {{\bf{v}}^{k - 1}}} \right\|_2^2\\
& \le \frac{{2n}}{{{\mu ^2}}} + \frac{2}{{{\mu ^2}\tau _2^2}}\left\| {{{\bf{v}}^k} - {{\bf{v}}^{k - 1}}} \right\|_2^2
\end{split}
\end{equation}
where the last inequality follows from $\| {\nabla {{\| {{{\bf{v}}^k}} \|}_{1,\varepsilon }}} \|_2^2 \le n$ when $\varepsilon > 0$.
Define ${\tilde{\bf{ z}}^k}: = ({{\bf{v}}^k},{{\bf{x}}^k},{{\bf{w}}^k},{{\bf{x}}^{k - 1}})$,
under the assumption that $\tilde {\mathcal{L}}({\tilde{\bf{ z}}^k})$ is lower semicontinuous, it is bounded from below.
Further, when (24) holds, $\tilde {\mathcal{L}}({{\bf{\tilde z}}^k})$ is nonincreasing by Lemma 3,
thus it is convergent. Then, form the definition of $\tilde {\mathcal{L}}$, we have
\begin{equation}
\begin{split}
&\tilde {\mathcal{L}}({{\tilde{\bf{ z}}}^1}) \ge \tilde {\mathcal{L}}({{\tilde{\bf{ z}}}^k})\\
& = \frac{1}{\mu }{\left\| {{{\bf{v}}^k}} \right\|_{1,\varepsilon }} + P({{\bf{x}}^k}) + \frac{\rho }{2}\left\| {{\bf{A}}{{\bf{x}}^k} - {\bf{y}} - {{\bf{v}}^k} - \frac{{{{\bf{w}}^k}}}{\rho }} \right\|_2^2\\
&~~~~~~~~~~~~~~~~~~~~~~~ - \frac{1}{{2\rho }}\left\| {{{\bf{w}}^k}} \right\|_2^2 + {c_2}\left\| {{{\bf{v}}^k} - {{\bf{v}}^{k - 1}}} \right\|_2^2\\
& \ge \frac{1}{\mu }{\left\| {{{\bf{v}}^k}} \right\|_{1,\varepsilon }} + P({{\bf{x}}^k}) + \frac{\rho }{2}\left\| {{\bf{A}}{{\bf{x}}^k} - {\bf{y}} - {{\bf{v}}^k} - \frac{{{{\bf{w}}^k}}}{\rho }} \right\|_2^2\\
&~~~~~~~~~~~~~~ - \frac{n}{{\rho {\mu ^2}}} + \left( {{c_2} - \frac{1}{{\rho {\mu ^2}\tau _2^2}}} \right)\left\| {{{\bf{v}}^k} - {{\bf{v}}^{k - 1}}} \right\|_2^2 \notag
\end{split}
\end{equation}
where the last inequality follows from (35). Since ${c_2} > \frac{1}{\rho {\mu ^2}\tau _2^2}$, when $P(\cdot)$ is coercive (e.g., for the hard-thresholding, soft-thresholding, SCAD, MC, and $\ell_q$-norm penalties), and by (35), it is easy to see that ${{\bf{v}}^k}$, ${{\bf{x}}^k}$ and ${{\bf{w}}^k}$ are bounded.

Since ${\tilde{\bf{ z}}^k}$ is bounded, there exists a convergent subsequence ${\tilde{\bf{ z}}^{{k_j}}}$ which converges to a cluster point ${\tilde{\bf{ z}}^*}$. Moreover, $\tilde {\mathcal{L}}({\tilde{\bf{ z}}^k})$ is convergent and $\tilde {\mathcal{L}}({\tilde{\bf{ z}}^k}) \ge \tilde {\mathcal{L}}({\tilde{\bf{ z}}^*})$ for any $k$ if ${c_3} > 0$. Then, it follows from Lemma 3 that
\begin{equation}
\begin{split}
&{c_{\rm{0}}}\sum\limits_{k = 1}^N {\left\| {{{\bf{x}}^{k + 1}} - {{\bf{x}}^k}} \right\|_2^2}  + {c_3}\sum\limits_{k = 1}^N {\left\| {{{\bf{v}}^{k + 1}} - {{\bf{v}}^k}} \right\|_2^2} \\
& \le \sum\limits_{k = 1}^N {\left[ {\tilde {\mathcal{L}}({{\tilde{\bf{ z}}}^k}) - \tilde {\mathcal{L}}({{\tilde{\bf{ z}}}^{k + 1}})} \right]} \\
& = \tilde {\mathcal{L}}({{\tilde{\bf{ z}}}^1}) - \tilde {\mathcal{L}}({{\tilde{\bf{ z}}}^{k + 1}})\\
& \le \tilde {\mathcal{L}}({{\tilde{\bf{ z}}}^1}) - \tilde {\mathcal{L}}({{\tilde{\bf{ z}}}^*}) < \infty. \notag
\end{split}
\end{equation}
Let $N \to \infty $, since ${c_{\rm{0}}} > 0$ and ${c_3} > 0$ when ${\tau _1} < 1/{\lambda _{\max }}({{\bf{A}}^T}{\bf{A}})$ and (24) are satisfied, we have
\begin{equation}
\begin{split}
&\sum\limits_{k = 1}^\infty  {\left\| {{{\bf{x}}^{k + 1}} - {{\bf{x}}^k}} \right\|_2^2}  < \infty\\
&\sum\limits_{k = 1}^\infty  {\left\| {{{\bf{v}}^{k + 1}} - {{\bf{v}}^k}} \right\|_2^2}  < \infty \notag
\end{split}
\end{equation}
which together with (33) implies
\begin{equation}
\sum\limits_{k = 1}^\infty  {\left\| {{{\bf{w}}^{k + 1}} - {{\bf{w}}^k}} \right\|_2^2}  < \infty .\notag
\end{equation}
Thus, we have $\mathop {\lim }\limits_{k \to \infty } \left\| {{{\bf{z}}^{k + 1}} - {{\bf{z}}^k}} \right\|_2^2 = 0$.

Next, we show that any cluster point of the sequence $\{ {{\bf{z}}^k}\}$ generated via (17), (23) and (16) is a stationary point of (21). From the optimality conditions, the sequence generated via (17), (23) and (16) satisfies
\begin{equation} 
\left\{ \begin{array}{l}
{\bf{0}} \in \partial {P({{\bf{x}}^{k + 1}})} - {{\bf{A}}^T}{{\bf{w}}^{k + 1}} + \rho {{\bf{A}}^T}({{\bf{v}}^{k + 1}} - {{\bf{v}}^k})\\
~~~~~~~~~~~~~~~~~~~~~~~~~~~~~~~~~~~ + \frac{\rho} {\tau _1}({{\bf{x}}^{k + 1}} - {{\bf{x}}^k}),\\
{\bf{0}} = \frac{1}{\mu} \nabla {\left\| {{{\bf{v}}^{k + 1}}} \right\|_{1,\varepsilon }} + {{\bf{w}}^{k + 1}} + \frac{1}{\mu {\tau _2}}({{\bf{v}}^{k + 1}} - {{\bf{v}}^k}),\\
{{\bf{w}}^{k + 1}} = {{\bf{w}}^k} - \rho ({\bf{A}}{{\bf{x}}^{k + 1}} - {\bf{y}} - {{\bf{v}}^{k + 1}}).
\end{array}\right.
\end{equation}
Let $\{ {{\bf{z}}^{{k_j}}}\}$ be a convergent subsequence of $\{ {{\bf{z}}^k}\}$,
since $\mathop {\lim }\limits_{k \to \infty } \left\| {{{\bf{z}}^{k + 1}} - {{\bf{z}}^k}} \right\|_2^2 = 0$,
${{\bf{z}}^{{k_j}}}$ and ${{\bf{z}}^{{k_j} + 1}}$ have the same limit point ${{\bf{z}}^*}: = ({{\bf{v}}^*},{{\bf{x}}^*},{{\bf{w}}^*})$.
Moreover, since $\tilde {\mathcal{L}}({\tilde{\bf{ z}}^k})$ is convergent, ${P({{\bf{x}}^{k}})}$ is also convergent.
Then, passing to the limit in (36) along the subsequence $\{ {{\bf{z}}^{{k_j}}}\}$ yields
\begin{equation}
{{\bf{A}}^T}{{\bf{w}}^*} \in \partial P({{\bf{x}}^*}),
~~ - {{\bf{w}}^*} = \frac{1}{\mu }\nabla {\left\| {{{\bf{v}}^*}} \right\|_{1,\varepsilon }},
~~ {\bf{A}}{{\bf{x}}^*} - {\bf{y}} = {{\bf{v}}^*}. \notag
\end{equation}
In particular, ${{\bf{z}}^*}$ is a stationary point of ${{\mathcal{L}}_\varepsilon}$.

\section{Proof of Lemma 5}
Let ${\tilde{\bf{ z}}^k}: = ({{\bf{v}}^k},{{\bf{x}}^k},{{\bf{w}}^k},{{\bf{x}}^{k - 1}})$,
from the definition of $\tilde {\mathcal{L}}({\tilde{\bf{ z}}^{k + 1}})$, we have
\begin{equation} 
{\partial _{\bf{x}}}\tilde {\mathcal{L}}({\tilde{\bf{ z}}^{k + 1}}) = \partial P({{\bf{x}}^{k + 1}}) - {{\bf{A}}^T}{{\bf{w}}^{k + 1}} + {{\bf{A}}^T}({{\bf{w}}^k} - {{\bf{w}}^{k + 1}}) \notag
\end{equation}
which together with the first relation in (36) yields
\begin{equation}
\begin{split}
\rho {{\bf{A}}^T}({{\bf{v}}^k} - {{\bf{v}}^{k + 1}}) &+ \frac{\rho }{{{\tau _1}}}({{\bf{x}}^k} - {{\bf{x}}^{k + 1}})\\
 &+ {{\bf{A}}^T}({{\bf{w}}^k} - {{\bf{w}}^{k + 1}}) \in {\partial _{\bf{x}}}\tilde {\mathcal{L}}({{\bf{\tilde z}}^{k + 1}}). \notag
\end{split}
\end{equation}
Moreover, we have
\begin{equation}
\begin{split}
&{\nabla _{\bf{v}}}\tilde {\mathcal{L}}({{\tilde{\bf{ z}}}^{k + 1}})\\
& = \frac{1}{\mu }\nabla {\left\| {{{\bf{v}}^{k + 1}}} \right\|_{1,\varepsilon }} + {{\bf{w}}^{k + 1}} \!-\! \rho ({{\bf{w}}^k} \!-\! {{\bf{w}}^{k + 1}}) \!+\! 2{c_2}({{\bf{v}}^{k + 1}} \!-\! {{\bf{v}}^k})\\
& = \rho ({{\bf{w}}^{k + 1}} - {{\bf{w}}^k}) + \left( {2{c_2} - \frac{1}{{\mu {\tau _2}}}} \right)({{\bf{v}}^{k + 1}} - {{\bf{v}}^k}) \notag
\end{split}
\end{equation}
where the second equality follows from the second relation in (36). Similarly,
\begin{equation}
\begin{split}
{\nabla _{\tilde{\bf{ v}}}}\tilde {\mathcal{L}}({\tilde{\bf{ z}}^{k + 1}}) &= 2{c_2}({{\bf{v}}^{k + 1}} - {{\bf{v}}^k}),\\
{\nabla _{\bf{w}}}\tilde {\mathcal{L}}({\tilde{\bf{ z}}^{k + 1}}) &= {\bf{A}}{{\bf{x}}^{k + 1}} - {\bf{y}} - {{\bf{v}}^{k + 1}} = \frac{1}{\rho }({{\bf{w}}^k} - {{\bf{w}}^{k + 1}}) . \notag
\end{split}
\end{equation}
Thus, we can find a constant ${c_5} > 0$ such that
\begin{equation}
\begin{split}
&{\rm{dist}}(0,\partial \tilde {\mathcal{L}}({{\tilde{\bf{ z}}}^{k{\rm{ + }}1}}))\\
& \le {c_5}({\left\| {{{\bf{x}}^{k + 1}} - {{\bf{x}}^k}} \right\|_2} + {\left\| {{{\bf{v}}^{k + 1}} - {{\bf{v}}^k}} \right\|_2} + {\left\| {{{\bf{w}}^k} - {{\bf{w}}^{k + 1}}} \right\|_2}) \notag
\end{split}
\end{equation}
which together with (33) consequently results in Lemma 5.

\section{Proof of Theorem 1}
Let ${{\bf{z}}^k}: = ({{\bf{v}}^k},{{\bf{x}}^k},{{\bf{w}}^k})$, based on the above lemmas, the rest proof of Theorem 1 is to show that the sequence $\{ {{\bf{z}}^k}\} $ has finite length, i.e.,
\begin{equation}
\sum\limits_{k = 0}^\infty  {{{\left\| {{{\bf{z}}^{k + 1}} - {{\bf{z}}^k}} \right\|}_{\rm{2}}}}  < \infty
\end{equation}
which implies that $\{ {{\bf{z}}^k}\}$ is a Cauchy sequence and thus is convergent.
Finally, the property (37) together with Lemma 4 implies that the sequence $\{ {{\bf{z}}^k}\}$ converges to a stationary point of ${{\mathcal{L}}_\varepsilon }$.
The derivation of (37) relies heavily on the Kurdyka-Lojasiewicz (KL) property of $\tilde {\mathcal{L}}$,
which holds if the penalty $P(\cdot)$ is a KL function.
This is the case of the hard-thresholding, soft-thresholding, SCAD, MC and $\ell_q$-norm penalties with $0 \le q \le 1$.
With the above lemmas, the proof of (37) follows similarly the proof of Theorem 3 in [55] with some minor changes, thus is omitted here for succinctness.

\end{document}